\documentclass[11pt,a4paper]{article}
\usepackage[square,numbers,sectionbib]{natbib}
\usepackage[margin=1in]{geometry}
\usepackage{times}
\usepackage{amsmath}
\usepackage{amsthm}
\usepackage{amssymb}
\usepackage{amsfonts}
\usepackage{enumerate}
\usepackage{subfigure}
\usepackage{graphicx}
\usepackage[font=small,labelfont=bf]{caption}
\usepackage{float}
\usepackage{dsfont}
\usepackage{url}
\usepackage[english]{babel}

\newcommand{\swallow}[1]{ }

\usepackage{graphicx}
\usepackage{url}
\usepackage{bm,array}

\usepackage{authblk}
\title{Alignment-Free Sequence Analysis and Applications}
\author[1]{Jie Ren}
\author[1,2]{Xin Bai}
\author[1]{Yang Young Lu}
\author[1]{Kujin Tang}
\author[3]{Ying Wang}
\author[4]{Gesine Reinert}
\author[1,2,*]{Fengzhu Sun}
\affil[1]{Molecular and Computational Biology Program, University of Southern California, Los Angeles, California, USA}
\affil[2]{Centre for Computational Systems Biology, School of Mathematical Sciences, Fudan University, Shanghai, China}
\affil[3]{Department of Automation, Xiamen University, Xiamen, Fujian, China}
\affil[4]{Department of Statistics, University of Oxford, Oxford, United Kingdom}
\affil[*]{Correspondence: fsun@usc.edu}
\date{}                     

\begin{document}

%

\maketitle


\begin{abstract}

Genome and metagenome comparisons based on large amounts of next generation sequencing (NGS) data pose significant challenges for alignment-based approaches due to the huge data size and the relatively short length of the reads. Alignment-free approaches based on the counts of word patterns in NGS data do not depend on the complete genome and are generally computationally efficient.  Thus, they contribute significantly to genome and metagenome comparison. Recently, novel statistical approaches have been developed for the comparison of both long and shotgun sequences. These approaches have been applied to many problems including the comparison of gene regulatory regions, genome sequences,  metagenomes, binning contigs in metagenomic data, identification of virus-host interactions, and detection of horizontal gene transfers.  We provide an updated review of these applications and other related developments of word-count based approaches for alignment-free sequence analysis.

\textbf{keywords}: sequence comparison, alignment, alignment-free, phylogeny, metagenomics, virus-host interaction, Markov chain, horizontal gene transfer
\end{abstract}




\section*{INTRODUCTION}
\label{intro}

Molecular sequence comparison is one of the most basic and fundamental problems in computational biology, and has been widely used to study the evolution of whole genome sequences and gene regulatory regions, gene function prediction, sequence assembly, and finding the relationships among microbial communities. The most widely used methods for molecular sequence comparison are alignment-based including the Smith-Waterman algorithm \cite{smith1981identification}, BLAST \cite{altschul1990basic}, BLAT\cite{kent2002blat}, etc. Although alignment-based approaches are most accurate and powerful for sequence comparison when they are feasible, their applications are limited in some situations. First, for whole genome comparison, there are many duplications, translocations, large insertions/delections, and horizontal gene transfers in the genomes.
This situation makes it difficult to use alignment-based methods to investigate the relationship among whole genome sequences.
Second, in the current next generation sequencing (NGS) era, investigators can sequence the genomes using NGS efficiently and economically. However, some parts of the genomes may not be sequenced due to the stochastic distribution of the reads along the genomes and the difficulties of sequencing some parts of the genomes, especially when the coverage is relatively low. Even if we can assemble the reads into long contigs, these contigs may not share long homologous regions making it challenging to study the relationships among the genomes using alignment in such situations.
Third, noncoding regions such as gene regulatory regions are not highly conserved except for some functional regions such as transcription binding sites, and cannot be reliably aligned.
Therefore, alignment-based approaches are not well suited to study the evolution of gene regulatory regions.
Fourth, alignment is not suitable to compare sequences of large divergence.
When we investigate the relationship between viruses and their hosts, infecting virus-host pairs may only share a tiny fraction of their genomes such as CRISPR regions, and thus alignment-based approaches can potentially identify the hosts of only a small fraction of viruses.
Fifth, many large genome and metagenome data sets from shotgun NGS sequencing are available and alignment-based methods are too time consuming. For all these scenarios, alignment-free methods for genome and metagenome comparison provide promising alternative approaches.

Alignment-free approaches for sequence comparison can be divided into several different groups: a) word-counts \cite{wang2009fungal,jun2010whole,sims2011whole,blaisdell1986measure,blaisdell1985markov,torney1990computation,wan2010alignment,reinert2009alignment,sims2009alignment,qi2004cvtree}, b) average longest common substrings \cite{ulitsky2006average}, shortest unique substrings \cite{yang2016estimator,yang2012alignment}, or a combination of both \cite{yang2013large}, c) sequence representation based  on chaos theory \cite{almeida2001analysis,wang2005spectrum,jeffrey1990chaos}, d) the moments of the positions of the nucleotides \cite{yau2008protein}, e) Fourier transformation \cite{yin2015improved}, f) information theory  \cite{vinga2013information}, and g) iterated maps \cite{almeida2013sequence}. Several excellent reviews on various alignment-free sequence comparison methods have been published \cite{zielezinski2017alignment,bonham2013alignment,song2014new,vinga2003alignment,li2010composition}

In this review, we concentrate on methods that can be applied to the comparison of sequences based on NGS data. Since the word-count-based approaches are the most adaptable to NGS reads data, we deal with word-count-based approaches as in \cite{song2014new}. These methods first count the number of occurrences of word patterns ($k$-mers, $k$-grams, $k$-tuples) along a sequence or in a NGS sample using different algorithms such as Jellyfish \cite{marccais2011fast}, DSK \cite{rizk2013dsk}, and KMC 2 \cite{deorowicz2015kmc}. Secondly, a similarity/dissimilarity measure is defined between any pair of sequences based on the word-count frequencies. Finally, various clustering algorithms such as hierarchical clustering and neighbor-joining are used to group the sequences. In the rest of the review, we use ``word" and ``$k$-mer" interchangeably.

The use of $k$-mer frequencies to compare molecular sequences traces back to the early work of Carl Woese and colleagues from the early  1970s to the mid 1980s when they generated oligonucleotide catalogs of 16S rRNA sequences from about 400 organisms \cite{sobieski198416s,fox1980phylogeny,fox1977classification,woese1985phylogenetic,mcgill1986characteristic,woese1985mycoplasmas}.
They used a similarity measure, $S_{AB}$, for two sequences $A$ and $B$  using $k$-mers similar to the Bray-Curtis dissimilarity \cite{fox1977comparative}.  When the whole 16S rRNA sequences for many organisms were available, they showed a positive correlation between the dissimilarity of two sequences using $k$-mers with the distance calculated by alignment although the correlation is not very high (0.40) \cite{woese1987bacterial}. Ragan et al. \cite{ragan2014molecular} gave an excellent review of these early efforts to study the relationships among 16S sequences using oligonucleotide patterns and compared the dendrograms derived using multiple sequence alignment, the similarity measure $S_{AB}$, and the newly developed $d_2^S$ statistic \cite{wan2010alignment,reinert2009alignment}. It was shown that the tree constructed based on $d_2^S$ for $k$ from 6 to 16 yielded the dendrogram that was most consistent with the maximum likelihood tree using multiple sequence alignment.

Many word-count-based methods for sequence comparison have been developed including the un-centered  correlation of word count vectors between two sequences \cite{torney1990computation}, $\chi^2$-statistics \cite{blaisdell1986measure,blaisdell1985markov}, composition vectors \cite{qi2004cvtree}, nucleotide relative abundances \cite{karlin1997compositional,karlin1995dinucleotide}, and the recently developed $d_2^*$ and $d_2^S$ statistics \cite{wan2010alignment,reinert2009alignment}.
It was shown that alignment-free methods are more robust than alignment-based methods especially against genetic rearrangements and horizontal gene transfers \cite{bernard2016alignment, chan2014inferring}. Since word frequencies are generally stable across different genomic regions, alignment-free methods work well even with sequences coming from different regions of the genomes.
Song et al. \cite{song2014new} presented an review of the  development and applications of these methods before 2013. In the current review, we provide further developments of $d_2^*$ and $d_2^S$ and their applications in recent years including a) how to determine the background Markov chain model of the sequences, b) genome, metagenome, and transcriptome comparison using Markov chains, c) inference of virus-bacterial host infectious associations, d) identification of horizontal gene transfers, and e) integrated software for alignment-free genome and metagenome comparison. We will also present an review of other developments related to $d_2^*$ and $d_2^S$ in recent years. For a recent review of other alignment-free sequence comparison methods and their applications, see \cite{zielezinski2017alignment}.

\section*{DETERMINATION OF THE BACKGROUND MARKOV CHAIN MODELS OF THE GENOMES}
\label{AFMC}

Alignment-free sequence comparison methods using $k$-mers generally involve counting the number of occurrences of words of length $k$ in genomic sequences and comparing sequences using dissimilarity measures defined based on $k$-mer frequencies. Different dissimilarity measures have been developed using a number of principles.
The measures can be broadly classified into two groups: measures that require background word frequencies and those that do not. Lu et al. \cite{lu2017cafe} developed a one-stop platform for computing a suite of 28 different alignment-free measures and provided various forms of visualization  tools including dendrograms, heatmaps, principal coordinate analysis and network display. The definitions of the 28 measures can be found in the supplementary material for \cite{lu2017cafe}.

For measures that do not require background word frequencies, the observed word frequency or word presence (or absence) are directly used to compute the dissimilarity measures. The measures include but are not limited to,
Euclidian distance ($Eu$),
Manhattan distance ($Ma$),
$d_2$ \cite{torney1990computation},
Feature Frequency Profiles ($FFP$) \cite{sims2009alignment},
Jensen-Shannon divergence ($JS$) \cite{narlikar2013one},
 Hamming distance, and Jaccard index. 
For measures that take background word frequency into account, dissimilarity between sequences is computed using the normalized word frequencies, where the expected word frequencies estimated using a background model are subtracted from the observed word frequencies to eliminate the background noise and enhance the signal.
This group of measures includes
$d_2^*$, $d_2^S$ \cite{reinert2009alignment, wan2010alignment} and their variants \cite{liu2011new,song2013alignment,ren2013multiple}, CVTree \cite{qi2004cvtree, qi2004whole}, Teeling \cite{teeling2004tetra}, EuF \cite{pride2006evidence}
and Willner \cite{karlin1997compositional,willner2009metagenomic}, where different forms of sequence background models are incorporated.

The second group of measures requires the knowledge about the approximate distribution of word counts in the background sequences.
Markov chains (MC) are widely used to model genomic sequences \cite{almagor1983markov} with many applications including
the study of dependencies between bases \citep{blaisdell1985markov},
the enrichment and depletion of certain word patterns \citep{pevzner1989linguistics},
prediction of occurrences of long word patterns from short patterns
\citep{hong1990prediction,arnold1988mono},
and the detection of signals in introns \citep{avery1987analysis}.
The defining feature of a MC model is the ``memorylessness'' property, that implies that the future state of the sequence can be well predicted solely based on its latest history without knowing the full history.
In particular, an $r$-th order MC assumes that the distribution of the future state only depends on the states of the past $r$ positions regardless of the earlier history, i.e. $P(X_t | X_{1} \dots X_{t-1}) = P(X_t | X_{t-r} \dots X_{t-1})$,
where $X_1, X_2,  \dots X_t, $ are the states in the sequence $X$, and $X_i$ takes its values from a finite alphabet of size $L$.
For DNA sequences, the alphabet set is ${\mathcal A}=\{A, C, G, T\}$.
The MC can be represented in the form of a $L^r \times L$ matrix, where the element in the matrix corresponds to the transition probability $P(w | w_1 w_2 \dots w_r)$, $ w \in \mathcal{A}$.
A 0-th order MC is the simplest case; in this case the positions in the sequence are \textit{independent and identically distributed (i.i.d.)}.

\subsection*{INFERENCE OF MC PROPERTIES FOR A LONG GENOMIC SEQUENCE}

For a long genomic sequence, efficient statistics are available to determine the order of the MC \cite{hoel1954test, anderson1957statistical, avery1999fitting, billingsley1961statisticalb,billingsley1961statisticala}.
For reviews on the application of MCs to molecular sequence analysis, see \cite{waterman1995introduction,reinert2000probabilistic,reinert2005statistics,ewens2005statistical}.
In particular,
under the hypothesis that the long sequence follows a $(k-2)$-th order MC,
it holds that twice the log-likelihood ratio of the likelihood of the sequence
under a $(k-1)$-th order MC versus that under the $(k-2)$-th order
MC model follows approximately a $\chi^2$-distribution with
${df}_k=(L-1)^2 L^{k-2}$ degrees of freedom.
The log-likelihood ratio can be approximated by the Pearson-type statistic
 \begin{equation}
  S_k = \sum_{\mathbf{w} \in \mathcal{A}^k } \frac{ \left(  N_{\mathbf{w}} - E_{\mathbf{w}} \right)^2 }{  E_{\mathbf{w}} },
  \label{S2_k equation}
  \end{equation}
where $\mathbf{w} = {w}_1 {w}_2\cdots {w}_k$
denotes a $k$-mer consisting of letters $w_i \in {\mathcal{A}}$,
$^{-}\mathbf{w} = w_2\cdots {w}_k$, $\mathbf{w}^{-} = {w}_1
{w}_2\cdots {w}_{k-1}$, and $^-\mathbf{w}^- = {w}_2 \cdots
{w}_{k-1}$,
$ N_{\mathbf{w}}$ denotes the count of the word
$\mathbf{w}$ in the sequence,
and $E_{\mathbf{w}} = \frac{N_{^{-}\mathbf{w} } N_{\mathbf{w}^{-}}}{N_{^{-}\mathbf{w}^{-}}}$
is the estimated
expected count of ${\mathbf{w}}$ if the sequence is generated by a
MC of order $(k-2)$, for $k \ge 3$.
For $k=2$, $N_{^{-}\mathbf{w}^{-}}$ is replaced by the total number of bases in the sequence.


Several estimators for the order of MC have been proposed based on the above results of the hypothesis testing.
Menéndez et al. \cite{menendez2011testing} proposed a procedure for estimating the order by performing a sequence of tests for increasing orders until the null hypothesis is accepted.
Papapetrou and Kugiumatzis \cite{papapetrou2013markov} similarly used sequential hypothesis tests to find the optimal order of MC based on the significance of the conditional mutual information (CMI) of different orders.
Moray and Weiss \cite{morvai2005order}, Peres and Shields \cite{peres2005two} and Dalevi et al. \cite{dalevi2006new} developed methods to estimate the order of a MC based on the observation of a maximal sharp transition of $|N_\mathbf{w} - E_\mathbf{w}|$ at the true order. Baigorri et al. \cite{baigorri2009markov} estimated the order of MC by considering the change of $\chi^2$-divergence involving $S_k$.
For the cases where a $\chi^2$-test fails due to inefficient data, Besag and Mondal \cite{besag2013exact} provided exact goodness-of-fit tests for Markov chains.

Model selection approaches have also been widely used in the determination of the order of MC.
The Akaike information criterion (AIC) \cite{tong1975determination}, AICc \cite{hurvich1995model}, the Bayesian information criterion (BIC) \cite{zhao2001determination}, and the Efficient Determination Criterion (EDC) \cite{dorea2006convergence} were proposed to estimate the order of MC, and their consistency were studied in Katz \cite{katz1981some} and Peres and Shields \cite{peres2005two}.
All of these model selection methods were formulated using the logarithm of the maximum likelihood of the sequence and a penalty term related to the number of parameters in the model.
Let $X$ be a sequence under the $r$-th order Markov model $\mathcal{M}_r$.
Then the log-maximum likelihood of the data under the model $\mathcal{M}_r$ is
\begin{align*}
l (X; \mathcal{M}_r) = \sum_{w_1 w_2 \dots w_r \in \mathcal{A}^r, w \in \mathcal{A}} N_{w_1 w_2 \dots w_r w} \log( \hat{P}(w | w_1 w_2 \dots w_r)),
\end{align*}
where $\hat{P}(w | w_1 w_2 \dots w_r)=\frac{N_{w_1 w_2 \dots w_r w}}{N_{w_1 w_2 \dots w_r}}$ is the estimated transition probability.
Then the optimal order $r^*$ of the MC is found by minimizing various criteria as follows.
\begin{align*}
AIC(r) &= -2 l (X; \mathcal{M}_r) + 2 |\mathcal{M}_r|, \\
AICc(r) &= AIC(r) + 2|\mathcal{M}_r| ( |\mathcal{M}_r| + 1)/( |X_r| - |\mathcal{M}_r| - 1), \\
BIC(r) &= -2 l (X; \mathcal{M}_r) + |\mathcal{M}_r| \log |X_r|, \\
EDC(r) &= -2 l (X; \mathcal{M}_r) + |\mathcal{M}_r| c(|X_r|),
\end{align*}
where $|X_r|$ is the data size of $X_r$, i.e. the total number of $(r+1)$-words in the sequence, $|\mathcal{M}_r|$ is the number of parameters in the model ($L^r \times L$ in this case), and $c(\cdot)$ is a general increasing function.
Narlikar et al. \cite{narlikar2013one} evaluated the AIC, AICc and BIC methods for estimating the order of a MC of a genomic sequence.
The results showed that the order of a MC had marked effects on the performance of sequence clustering and classifications.
The MC order obtained based on the BIC optimality criterion yielded the best performance among all the model selection criteria.

\subsection*{INFERENCE OF MC PROPERTIES BASED ON NGS DATA}

One successful application of alignment-free methods is comparing different genomes using NGS reads data for which each sample contains millions of short reads randomly sampled from different parts of the genomes.
For NGS reads data, it is challenging to assemble short reads to recover the original genomic sequences.
Ren et al. \cite{ren2016inference} developed an assembly-free method to estimate background MCs based solely on short reads.
The NGS reads data are modeled as generated by a two-layer stochastic process: first, a (un-observed) long MC sequence is generated, and second, short reads are randomly sampled from the long MC sequence.

The classic statistic $S_k$ defined in equation \ref{S2_k equation} for the long sequence was extended to $S_k^R$ for the NGS data by replacing the word frequencies in a long sequence with that in NGS short reads.
Let $N_\mathbf{w}^R$ be the count of the $k$-word $\mathbf{w}$ in the NGS short reads (the superscript $R$ refers to the ``read'' data). Define
\begin{align}
S_k^R = \sum_{\mathbf{w} \in \mathcal{A}^k } \frac{ \left(  N_{\mathbf{w}}^R - E_{\mathbf{w}}^R \right)^2 }{  E_{\mathbf{w}}^R }.
\end{align}
Due to the additional randomness introduced in the process of sampling short reads from genomic sequences, the new statistic $S_k^R$ no longer follows the classic $\chi^2$- distribution.
Instead, it was shown that $S_k^R$ follows a gamma distribution when the reads are sampled based on the Lander-Waterman model \cite{lander1988genomic}.
In particular,
let $f_i$ be the fraction of the genome that is covered by exactly $i$ reads, $i=1,2,\cdots$. Define the effective coverage
\begin{align}
d = \frac{\sum_{i} i^2 f_i}{\sum_{i} i f_i}.
\end{align}
The statistic $S_k^R/d$ has an approximate $\chi^2$-distribution
    with ${df}_k = (L-1)^2 L^{k-2}$ degrees-of-freedom; equivalently, the
    statistic $S_k^R$ has an approximate gamma distribution with shape
    parameter ${df}_k/2$ and scale parameter $2d$.
Several estimators for the order of a MC based on NGS data using various criteria, such as the sharp transition of $S_k^R$, AIC and BIC, were proposed and compared in \cite{ren2016inference}, by extending the classical order estimators for long genomic sequences to those for NGS data.

\section*{APPLICATIONS OF THE ALIGNMENT-FREE METHODS TO COMPARATIVE GENOMICS}

Among the various alignment-free sequence comparison methods, the measures using normalized $k$-mer counts, $d_2^*$ and $d_2^S$ \cite{wan2010alignment,reinert2009alignment,song2014new}, have been shown to have
superior performance for comparing genomic sequences.
Wan et al. \cite{wan2010alignment} and Burden et al. \cite{burden2012alignment} studied the theoretical statistical properties of the $d_2^*$ and $d_2^S$ measures.
Song et al. \cite{song2013alignment} extended the definition of $d_2^*$ and $d_2^S$ from two long genomic sequences to the comparison of two samples based on NGS reads data, and investigated theoretically the properties of the measures.
As an application, the relationship of 13 tropical tree species in \cite{cannon2010assembly} were revealed without assembly using $d_2^*$ and $d_2^S$.
Ren et al. \cite{ren2016inference} clustered genomic sequences of 28 vertebrate species based on NGS reads using $d_2^*$ and $d_2^S$ under different MC models.
Using the appropriate order of MC, the pairwise dissimilarity scores using $d_2^*$ and $d_2^S$ are highly correlated (Spearman's rank correlation coefficient 0.92) with the true pairwise evolutionary distances inferred based on the multiple sequence alignment of homologous genes in \cite{miller200728}.
Compared to $d_2^*$, $d_2^S$ is less affected by the order of the MC model. For example, the Spearman's rank correlation coefficient using $d_2^S$  is 0.86 even under the i.i.d model.

Bernard et al. \cite{bernard2016alignment} and  Chan et al. \cite{chan2014inferring}  systematically assessed the performance of various alignment-free measures under different evolutionary scenarios using simulations and empirical data.
The results showed that the alignment-free methods are sensitive to sequence divergence, less sensitive to lateral genetic transfer, and robust against genome rearrangement, among-site rate heterogeneity and compositional biases.
Chan et al. \cite{chan2014inferring} performed phylogenetic inference using alignment-free measures for 4,156 nucleotide sequences.
The topology obtained using $d_2^S$ was most congruent with the phylogeny inferred using multiple sequence alignment.
Similarly, the relationship among 143 bacteria and archaea genomes \cite{bernard2016alignment, bernard2016recapitulating}, 63 Enterobacteriaceae genomes \cite{yi2013co}, 27 Escheriachia coli and Shigella genomes \cite{bernard2016alignment, yi2013co}, 21 primate genomes \cite{lu2017cafe} and 27 primate mitochondrial genomes \cite{leimeister2014fast}, 14 plants \cite{leimeister2014fast}, and 8 Yersinia genomes \cite{bernard2016alignment} were inferred using $d_2^S$ and compared with the evolutionary tree built based on multiple sequence alignment. Despite some incongruence, the clustering results in general had highly similar structures with the classical evolutionary trees.

To evaluate the robustness of the clustering, different resampling methods, including jackknife \cite{bernard2016alignment} and bootstrap \cite{ren2016inference,fan2015assembly}, were applied for resampling sequences to provide a measure of robustness for the branches in the inferred clustering tree.
The studies showed that alignment-free methods can accurately recover phylogenetic relationship even with low sequencing coverage.
The time complexity for alignment-free methods was significantly lower compared to the traditional maximum likelihood and Bayesian methods based on multiple sequence alignment \cite{fan2015assembly}.
It was estimated that alignment-free methods are approximately 140-fold faster than the traditional methods \cite{chan2014inferring}.
Normalization of the background and including inexact matches increases the time complexity.
Alignment-free methods based on $k$-mers lend themselves to parallel algorithms, and parallel computational methods have been applied to achieve speedup and scalability for alignment-free methods \cite{cattaneo2017effective}.
When $k$ is large, memory is a main limitation for storing $k$-mer counts and computing alignment-free measures \cite{bernard2017alignment}.

\section*{PREDICTION OF VIRUS-PROKARYOTIC HOST INTERACTIONS USING ALIGNMENT-FREE METHODS}

It is widely recognized that bacteria and archaea (prokaryotes) play important roles in many ecosystems and significantly impact the health of humans, animals, and plants \cite{rappe2003uncultured}.
However, much less is known about the viruses that infect prokaryotes.
Since viral infections can lead to lysis of host cells, viruses consequently can indirectly impact ecological processes by regulating and controlling the abundance of prokaryotes.
Metagenomic sequencing, that uses NGS to recover genetic material of microbial organisms from environment samples, can be used for high-throughput identification of bacteria, archaea, and viruses regardless of culturability.
Increasing numbers of new viruses have been discovered by assembling short reads from various environments including human gut \cite{dutilh2014highly, norman2015disease, reyes2015gut, minot2011human, waller2014classification}, ocean \cite{brum2015patterns, reyes2010viruses, zhang2005rna}, and soil \cite{pearce2012metagenomic, adriaenssens2015metagenomic, zablocki2014high}.
Yet, their biological functions and prokaryotic hosts cannot be directly inferred from the metagenomic data.

A few computational approaches have been developed recently for predicting the host given a viral sequence.
The most straight forward method is alignment-based gene homology search and CRISPR search between virus and host genomes \cite{roux2015virsorter}.
However, not many viruses share regions with hosts and not many hosts have CRISPR spacers.
In contrast, alignment-free methods can be powerful for revealing virus-host interaction relationships, because it is observed that viruses share highly similar $k$-mer usage with their hosts, possibly due to the fact that virus replication is dependent on translational machinery of its host \cite{pride2006evidence}.
Edwards et al. \cite{edwards2016computational} and Roux et al. \cite{roux2015viral} used Euclidean and Manhattan distances based on tetramers ($k=4$) to measure the distance between viruses and hosts, and predicted the host as the one with the smallest distance to the query virus.

\begin{figure}[ht]
\centering
\includegraphics[scale=0.65]{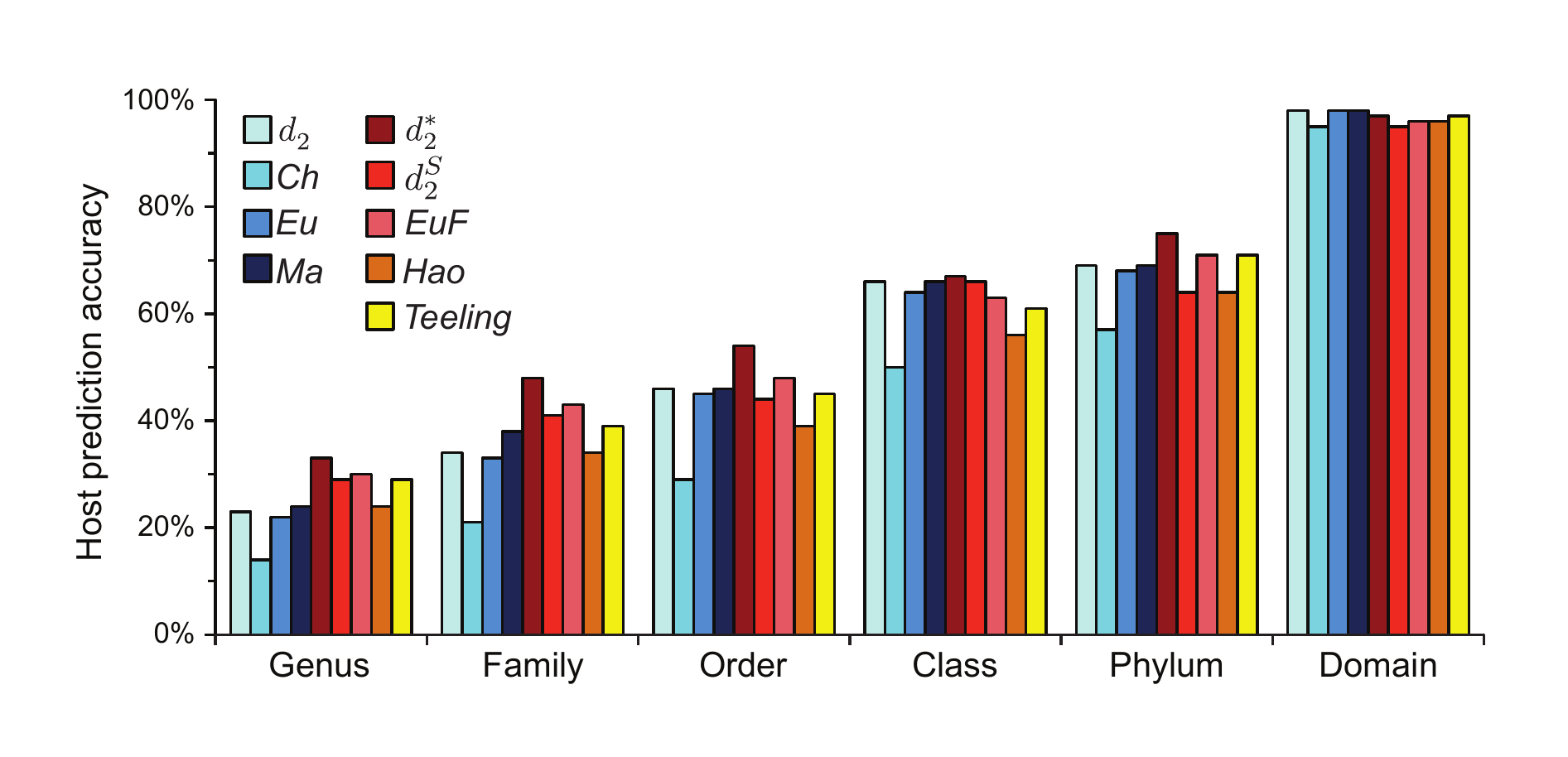}
\caption[Prediction accuracy using ONF with various distance/dissimilarity measures at $k$-mer length 6 on the benchmark dataset]{Prediction accuracy using various distance/dissimilarity measures at $k$-mer length 6 on a benchmark data set of 1,427 complete viral RefSeq genomes whose hosts are known versus $\sim$ 32,000 possible archaea and bacteria host genomes. Predictions were made for all 1,427 viruses from Ahlgren et al. \cite{ahlgren2017alignment}.}
\label{vhm_fig2}
\end{figure}

Ahlgren et al. \cite{ahlgren2017alignment} conducted a comprehensive evaluation of alignment-free dissimilarity measures over various $k$-mer lengths for host prediction.
The study evaluated a suite of 11 measures including those based on the observed word frequencies such as Euclidean and Manhattan distances and those based on normalized word frequencies such as $d_2^*$ and $d_2^S$.
The prediction accuracy of the measures were assessed based on the largest benchmark dataset containing 1,427 virus isolate genomes whose true hosts are known and $\sim$32,000 prokaryotic genomes as host candidates.
The measures based on normalized frequencies in general have better discriminatory power of separating true interacting virus-host pairs from random pairs than those based on observed word frequencies.
Increasing $k$-mer length from 4 to 6 also improves the discriminatory power.
Among the 11 measures, $d_2^*$ at $k= 6$ and a second order MC yielded the highest host prediction accuracy (\textbf{Figure \ref{vhm_fig2}}).
Requiring a minimum dissimilarity score for making predictions (thresholding) and taking the consensus
of the 30 most similar hosts further improved accuracy.
While prediction accuracy decreases for shorter contigs,
the method is able to make decent predictions on contigs as short as 5 kbp.
A software called VirHostMatcher was developed for predicting hosts of viruses and visualizing the predicted results using alignment-free methods.

Following the same principle that the virus-host genomes tend to have high similarity, Galiez et al. \cite{galiez2017wish} developed a program, WIsH, that computes the likelihood of the query viral sequence under each of the Markov models for candidate bacteria genomes, and predicts the host as the one whose model yields the highest likelihood.
Since the program only relies on the Markov models for bacteria complete genomes, the method achieves decent accuracy even for viral contigs as short as 3 kbp, and it is generally faster than VirHostMatcher.
WIsH uses a fixed 8th order MC to model the bacteria genomes,
so the method may not be readily applicable for metagenomic contigs where the host contigs are so short that no sufficient data is available for estimating a high order MC.

Another group of host prediction methods is based on the observation that similar viruses often share the same host range.
Different virus-virus similarity measures have been investigated using various principles \cite{paez2016uncovering, lima2008reticulate, shapiro2017gene}, and the clusters in the gene-based virus-virus similarity network show high association with the host classes \cite{shapiro2017gene}.
Villarroel et al. \cite{villarroel2016hostphinder} developed a host prediction tool, HostPhinder, that predicts the host of a query virus as the host of the most similar reference virus. The similarity was defined based on the proportion of the shared $k$-mers between the query and the reference virus genomes.
Zhang et al. \cite{zhang2017prediction} developed machine learning based classifiers to predict if a query virus can infect a particular host genus, based on the common $k$-mer features learned from the existing infectious viruses.
However, the method is only applicable to hosts that have a relatively large number of known infecting viruses.

\section*{GENOME AND TRANSCRIPTOME COMPARISON USING ALIGNMENT-FREE APPROACHES WITH VARIABLE LENGTH MARKOV CHAINS}

Using Fixed Order Markov Chains (FOMC) to model the background sequence has several potential limitations.  First, the MC order needs to be set manually. However, for most sequences of interest, there is no prior knowledge available for setting the correct MC order. Second, FOMC is not structurally rich. The number of parameters in an $r$-th order MC is $(L-1) L^r$ where $L$ is the alphabet size, and there are no MC models with number of parameters between $(L-1) L^r$ and $(L-1) L^{r+1}$.
Third, the number of parameters grows exponentially with the  MC order \textit{r}. When the length of the sequence is short or sequencing depth is relatively low, the parameters cannot be accurately estimated.

Therefore, Liao et al. \cite{liao2016alignment} investigated the use of  the data-driven Variable Length Markov Chain (VLMC) \cite{buhlmann1999variable} model as an alternative to FOMC to model background sequences. VLMC was originally designed for modeling one long sequence and was represented as a context tree structure \cite{buhlmann1999variable, rissanen1983universal}. Liao et al. \cite{liao2016alignment} designed a three-step approach for prunning a tree based on NGS short reads data. First, a full prefix tree  based on $1, 2, {\cdots}, \textit{10}$-mer frequency vectors was built. However, the tree usually overfits the data.
Second, the full prefix tree was pruned to remove the redundant branches based on the Kullback-Leibler divergence \cite{kullback1951information}. The pruned tree is called a context tree \cite{rissanen1983universal}. The threshold value \textit{K} for the Kullback-Leibler divergence determines the complexity of the pruned tree. The value of \textit{K} was chosen by optimizing the Akaike Information Criterion (AIC) \cite{akaike1987factor} designed for the high-throughput sequencing data. AIC measures the relative quality of statistical models for a given set of data. Third, transition probabilities were estimated with respect to the VLMC from the context tree, and the probabilities of words were then computed accordingly.

Liao et al. \cite{liao2016alignment} evaluated the performance of  $d^S_2$ and $d^*_{2}$ using both simulations and real data. It was shown that VLMC outperformed FOMC to model the background sequences in transcriptomic and metatranscriptomic samples.
Moreover, $d^S_2 $ based on VLMC background model can identify underlying relationships among metatranscriptomic samples from different microbial communities, and can reveal a gradient relationship among the metatranscriptomic samples. VLMC is easier to apply than FOMC because of being free from MC order selections. The flexible number of parameters in VLMC avoids estimating the vast number of parameters of high-order MC under limited sequencing depth. 
In contrast, the VLMC model does not work as well as FOMC for investigating the relationship among whole genome or metagenome data. It was hypothesized that whole genomes and metagenomes contain mixtures of coding and noncoding regions and are too complex to be modeled by relatively concise VLMC models.
Yet, the coding regions are more homogeneous than the whole genome.
The clustering performance can be improved for metatranscriptomic data using the VLMC to model the background sequence, but not for whole genome or metagenomic data.
For the comparison of metagenomes, Jiang et al. \cite{jiang2012comparison} showed that  $d_2^S$ with the i.i.d background model and $k$-mer length between 6 to 9 bps generally performs well compared to other measures.

It is time-consuming to model VLMC due to the generation and the pruning of the prefix tree. Behnam and Smith \cite{behnam2014amordad} measured the dissimilarity between metagenomic samples with dot product distance based on the i.i.d. model, and they integrated a randomized hashing strategy based on locality-sensitive hashing and the regular nearest neighbor graph to reach logarithmic query time for identifying similar metagenomes even as the database size reaches into the millions. Meanwhile, also focusing on fast comparisons among large-scale multiple metagenomic samples, Benoit et al. \cite{benoit2016multiple} developed the program,  Simka, to compute 16 standard ecological distances by a parallel \textit{k}-mer counting strategy on multiple data sets. Simka was able to compute in a few hours both qualitative and quantitative ecological distances based on hundreds of metagenomic samples.

\section*{IMPROVING METAGENOMIC CONTIG BINNING USING $d_2^S$ }

 Wang et al. \cite{wang2017improving} used $d_2^S$ to improve contig binning. Assigning assembled contigs into discrete clusters, known as bins, is a key step toward investigating the taxonomic structure of microbial communities \cite{mande2012classification}. Contig binning using \textit{k}-mer composition is based on the observation that relative sequence compositions are similar across different regions of the same genome, but differ between distinct genomes \cite{karlin1997compositional,dick2009community}. Contigs in the same bin are expected to come from the same taxonomic group.  Three different types of strategies have been used to bin contigs: sequence composition, abundance and a hybrid between the two.  Sequence composition based methods use $k$-mer frequencies with $k\mathrm{=}$2-6 as genomic signatures of contigs \cite{leung2011robust, kislyuk2009unsupervised}. Abundance based methods use the relative abundance levels of species and the distribution of the number of reads containing certain $k$-mers to bin contigs  \cite{wu2011novel,wang2015mbbc}. The hybrid approaches use both composition and abundance of $k$-mers to bin contigs \cite{wu2014maxbin,lin2016accurate}. Most of the currently available binning methods used the frequency of \textit{k}-mers directly, but this represented absolute, not relative, sequence composition. Here, ``absolute" frequency refers to the number of occurrences of a \textit{k}-mer over the total number of occurrences of all \textit{k}-mers. On the other hand, ``relative" frequency refers to the difference between the observed frequency of a \textit{k}-mer and the corresponding expected frequency under a given background model.
The dissimilarity measures $d^S_2$ based on relative frequencies of \textit{k}-mers have been successfully used for sequence comparison as reviewed above.
Therefore, we expected that calculating the dissimilarity between contigs using $d_2^S$ would improve contig binning compared to other contig binning methods that are based on the difference of absolute $k$-mer frequencies. However, directly using $d_2^S$ for contig binning is too time consuming and is impractical for most metagenomic data.

Instead of binning contigs directly using $d_2^S$, Wang et al. \cite{wang2017improving} developed $d^S_2\mathrm{Bin}$ that uses $d_2^S$ to improve reasonable contig binning results using other fast and efficient programs such as MetaCluster3.0 \cite{leung2011robust}, MetaWatt \cite{strous2012binning}, SCIMM \cite{kelley2010clustering}, MaxBin1.0 \cite{wu2014maxbin}, and MyCC \cite{lin2016accurate}. Each contig was modeled with a MC based on its \textit{k}-mer frequency vector. The center of the bin was represented by the average \textit{k}-mer frequency vectors of all contigs in this bin and was also modeled with a MC. Then, $d^S_2$ was used to measure the dissimilarity between a contig and the center of a bin based on relative $k$-mer composition. Finally, a K-means clustering algorithm was applied to cluster the contigs based on the $d^S_2$ dissimilarities.  \textit{Recall}, \textit{precision} and Adjusted Rand Index (\textit{ARI}) were used to evaluate the binning performance. Wang et al. \cite{wang2017improving} showed that $d^S_2\mathrm{Bin}$ consistently achieved the best performance with $k\mathrm{=6}$-mers under the i.i.d. background model.
$d^S_2\mathrm{Bin}$ improves the binning performance in 28 out of 30 testing experiments.
Experiments showed that $d_2^S$ accurately measures the dissimilarity between contigs of metagenomic reads and that measures defined based on relative sequence composition are more suitable for contig binning.
Also, $d^S_2\mathrm{Bin}$ can be applied to any existing contig-binning tools for single metagenomic samples to improve binning results.

\section*{IMPROVING THE IDENTIFICATION OF HORIZONTAL GENE TRANSFER USING $d_2^*$ OR CVTree}

Horizontal gene transfer (HGT) or lateral gene transfer (LGT) describe the transmission of genetic material between organisms that are not in a parent-offspring relationship.
HGT plays an important role in the evolution of microbes and is responsible for metabolic adaption \cite{pal2005adaptive} and the spread of antibiotic resistance \cite{gyles2014horizontally}. Existing computational methods for HGT inference can be broadly separated into two groups: alignment-based and alignment-free methods.

Alignment-based, or phylogenetic methods for detecting HGT rely on phylogenetic conflicts; that is, finding genes whose phylogenetic relationships among multiple organisms differ significantly from that of other genes \cite{ravenhall2015inferring,lu2016computational}. Although alignment-based methods are considered to be the gold standard \cite{keeling2008horizontal} for HGT detection because of their explicit models, finding topological incongruences is computationally demanding, requires large memory, and requires that genomes of interest are annotated and their phylogenetic relationships are known.
In addition, alignment-based methods can only be applied to coding sequences and thus have limited ability to detect horizontal transfer in non-coding regions.

Instead, alignment-free methods, also called compositional parametric methods, can be used to avoid these limitations. Alignment-free methods infer horizontal gene transfer by detection of regions in a genome with atypical word pattern composition based on the observation that sequences transferred from donor genomes have different composition signatures from that of the host genome \cite{karlin1995dinucleotide}. Recently, Cong et al. \cite{cong2017robust,cong2016novel,cong2016exploring} introduced TF-IDF as a scalable alignment-free approach for HGT detection in large molecular-sequence data sets by combining multiple genomes and $k$-mer frequencies. However, these methods require the phylogenetic relationship among a group of genomes and they can only detect HGT within this group of genomes. More widely used alignment-free methods apply a sliding window approach to scan a single genome and calculate the dissimilarity between each window and the whole genome. Consecutive windows with dissimilarity higher than a threshold are inferred as HGT. The performances of $k$-mer-based alignment-free methods depend largely on the choice of dissimilarity measures between a genomic region and the whole genome, on the $k$-mer length, on the sliding window size, and on the evolutionary distance between host and donor genomes. Manhattan and Euclidean distances between the $k$-mer frequency vector of a genomic region and that of the whole genome are the most frequently used measures for detecting HGTs because of their simplicity. For example, Dufraigne et al. \cite{dufraigne2005detection} analyzed HGT regions of 22 genomes by using Euclidean distance with $k$-mer length of 4 bps.
Rajan et al. \cite{rajan2007identification} used Manhattan distance with $k$-mer length of 5 bps to detect HGT in 50 diverse bacterial genomes.

Several papers compared the performances of different dissimilarity measures for HGT detection. Because the true HGT history is unknown, the evaluation and benchmarking of HGT detection methods typically relies on simulated artificial genomes, for which the true simulated history is known.
Tsirigos and Rigoutsos \cite{tsirigos2005new}  investigated several dissimilarity measures between the relative frequencies of a genomic region and the whole genome under the i.i.d. model including correlation, covariance, Manhattan distance, Mahalanobis distance, and Kullback--Leibler (KL) distance for HGT detection. They showed that $k$-mers of length 6-8 bps with covariance dissimilarity perform the best under their simulated situations.
Becq et al. \cite{becq2010benchmark} reviewed alignment-free methods on horizontal gene transfer detection and showed that $k$-mer-based methods with a 5 kbps sliding window outperformed other alignment-free methods based on features such as GC content \cite{karlin2001detecting}, codon usage \cite{karlin2001detecting} and dinucleotide content \cite{karlin1995dinucleotide}. However, they only tested Euclidean distance with $k$-mer length 4 bps as genomic signature \cite{dufraigne2005detection} for $k$-mer-based methods.  

\begin{figure}[htbp]
\centering\includegraphics[width=5.0in]{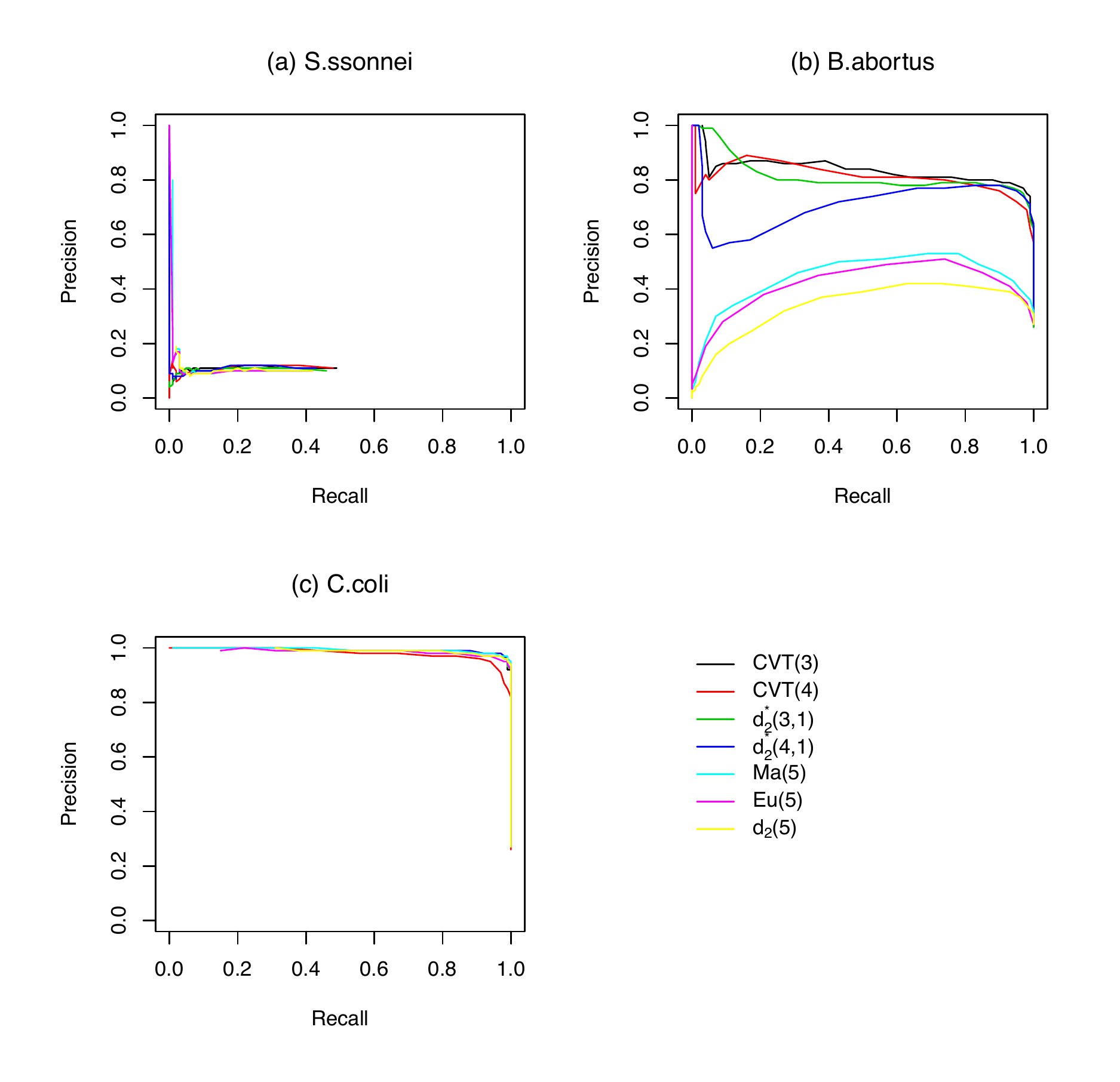}
\caption{The Precision-Recall Curves (PRC) of different HGT detection methods along artificial genomes using \textit{E.coli} as host genome. Precision and recall values were calculated by defining different thresholds for HGT. Numbers in the brackets indicate the word length $k$ used by different methods and Markov order used by $d_2^*$.  For example, $d_2^*(3, 1)$ means that $d_2^*$ was the dissimilarity measure with word length 3 and Markov order 1. (a) PRC when using \textit{S.sonnei\emph{}} as donor genome, which is at the same species level as \textit{E.coli}. None of the methods perform well. (b) PRC when using \textit{B.abortus} as donor genome, which is at the same class but different order level as \textit{E.coli}. In this scenario, $CVT(3)$, $CVT(4)$, $d_2^*(3,1)$, and $d_2^*(4,1)$ outperform other methods. (c) PRC when using \textit{C.coli} as donor genome, which has different order level from \textit{E.coli}. All methods perform reasonably well.}\label{fig:HGT}
\end{figure}

Recently, we evaluated the performance of different dissimilarity measures including Manhattan, Euclidean,  CVtree, $d_2$, $d_2^*$, $d_2^s$ with different choices of $k$-mer length and Markov order. We also studied the influence of window size and evolutionary distance between host and donor genomes on HGT detection by both simulation and real data in terms of \swallow{$F_1$ score and} precision-recall curve (PRC).
We showed that none of these dissimilarity measures work well when the donor and host genomes are within the same order level since the donor and host genomes are too similar and it is challenging to distinguish the transferred regions.
All dissimilarity measures perform well when the donor and host genomes are in different class levels since the host and donor genomes are highly different and most of these methods can identify their differences.
For HGT between genomes  from different order levels but in the same class level, background adjusted dissimilarity measures that consider Markov order of sequences, such as CVtree with $k$ = 4 and $d_2^*$ with $k$ = 3 and Markov order 1 can achieve significantly better performance than the other methods. The PRC results for different scenarios are shown in \textbf{Figure \ref{fig:HGT}}.

Therefore, $k$-mer-based alignment-free methods for HGT detection are suitable when host and donor genomes are in different order levels and HGT length is greater than 5 kbps.
Therefore, alignment-free methods should not replace alignment-based methods in all cases. Instead, they are complimentary as each has unique advantages in different scenarios and they also tend to find complimentary sets of HGT regions \cite{tamames2008estimating}. Alignment-free methods are preferred when no evolutionary trees are available or genomes are not well annotated. Our study suggests that CVTree with word length of 4, $d_2^*$ with word length 3, Markov order 1, and $d_2^*$ with word length 4, Markov order 1 all perform well in most situations.

%
%

\section*{OTHER WORD-COUNT BASED APPROACHES FOR SEQUENCE COMPARISON}

Many other sequence dissimilarity measures based on $k$-mer frequencies have been developed in recent years. 
Liu et al. \cite{liu2011new} proposed local alignment-free measures by summing up the maximal pairwise scores between any sub-fragments of a fixed length in the sequence.
Ren et al. \cite{ren2013multiple} developed a suite of alignment-free multiple sequence comparison methods to enable measuring similarity among a set of more than two sequences. Several alignment-free  methods incorporating potential mismatches, sequencing errors,  or spaced word patterns have been developed for sequence comparison
\cite{yi2013co, leimeister2014fast,goke2012estimation, horwege2014spaced}.
Fan et al. \cite{fan2015assembly} developed a method called Assembly and Alignment-Free (AFF) that defines the distance based on the proportion of shared $k$-mers as an indication of the amount of divergence between the species.

In most of the dissimilarity measures reviewed above,  the $k$-mers are treated equally. Differential weighting of the $k$-mers may help study the relationship among the sequences.
Patil and McHardy \cite{patil2013alignment} generalized Euclidean distance to a weighted Euclidean distance where the weights are learned from the training data, and evaluated on the independent test data. The learned weighted Euclidean distances specified for a group of species increase the accuracy for inferring taxonomic relationships of a new species from the same group.

Qian and Luan \cite{qian2017weighted} developed an alternative approach for weighting the different $k$-mers by maximizing the weighted $L_1$ norm between the frequency vectors among all the sequences with $c_\mathbf{w}$ being the weight for the word $\mathbf{w}$.
Qian and Luan \cite{qian2017weighted} proposed to maximize
\[  \sum_{\mathbf{w} \in \mathcal{A}^k} \sum_{i,j = 1}^n  c_\mathbf{w} | f_{i \mathbf{w}} - f_{j \mathbf{w}}|,  \]
with the constraint of  $ \sum_{\mathbf{w} \in  \mathcal{A}^k} c_\mathbf{w} = 1$,
where $n$ is the number of sequences to be compared. Once the values of $c_\mathbf{w}$ were determined,  they modified the definitions of $d_2, d_2^*$ and $d_2^S$ by putting the weight $c_\mathbf{w}$ in front of the corresponding terms. Applications to the identification of homologous genes and cis-regulatory modules (CRM) showed that the weighted versions of these measures outperformed the original ones.

 It was reasoned that if a $k$-mer is present/absent in a small fraction or most of the sequences, it does not markedly contribute to distinguishing the different sequences. Therefore, weighting the different $k$-mers according to the frequency of being present/absent in the sequences of interest can increase our understanding of the relationships among the sequences \cite{murray2017kwip}. For a $k$-mer $\mathbf{w}$, let $F_\mathbf{w}$ be the fraction of the sequences with $\mathbf{w}$ present. The entropy of the word is defined as
\[ H_\mathbf{w} = - ( F_\mathbf{w} \log_2 (F_\mathbf{w})  + (1 - F_\mathbf{w}) \log_2 (1  - F_\mathbf{w})). \]
The weighted similarity measure between sequence $i$ and sequence $j$ was defined as
\[ K_{ij} = \sum_{\mathbf{w} \in \emph{A}^k}  H_\mathbf{w} f_{i\mathbf{w}} f_{j\mathbf{w}}, \]
 and then normalized using
\[ K^{'}_{ij} = \frac{K_{ij}}{\sqrt{K_{ii} K_{jj}}}.  \]
Finally, the dissimilarity between the two sequences was defined as $d_{ij} = \sqrt{2(1 - K^{'}_{ij})}$ \cite{murray2017kwip}. To speed up computational time as well as to save memory, Murray et al. \cite{murray2017kwip} bin the $k$-mers into different groups so that a group contains multiple $k$-mers. The authors showed that this weighted version outperformed the traditional $d_2$ statistic and the Mash program \cite{ondov2016mash}.


\section*{DETERMINATION OF WORD SIZE $k$}
In alignment-free sequence comparison using word counts, an important yet challenging problem is the length of word patterns. Although many studies are available, there are still no definitive answers to the optimal choice of word length. The optimal word length depends on the statistical measures for comparing the sequences, and the background models, the lengths, and the diversity of the sequences to be compared. For example, if the sequences are short, the optimal word length may be short since the sequences do not contain a large number of distinct words. Otherwise, the sequences may rarely share common word patterns.
However, short word patterns do not have high power to discriminate closely related sequences. If the sequences to be compared are highly similar, we expect that the optimal word length should be long as short word patterns will not be able to distinguish them. On the other hand, if the sequences to be compared are diverse, relatively short word patterns  may suffice to distinguish the sequences.

Recently, Bai et al. \cite{bai2017optimal} investigated the optimal word length when comparing two Markovian sequences using the $\chi^2$-statistic in \cite{blaisdell1986measure}.
Bai et al. \cite{bai2017optimal} framed sequence comparison as a hypothesis testing problem of evaluating if the two sequences come from two different Markov chains and used power under the alternative hypothesis as an optimality criterion.  They showed both theoretically and by simulations  that the optimal word length equals the maximum of the Markov orders of the two sequences plus one. This conclusion also holds for NGS data. Using the estimated Markov orders resulted in minimal loss of power when comparing two sequences. Applications to real sequences to find homologs of the human protein HSLIPAS and the cis-regulatory modules (CRM) in four mouse tissues (forebrain, heart, limb and midbrain) confirmed the theoretical results. Preliminary simulation results showed that this $k$-mer length may also be optimal for other measures including  CVTree \cite{qi2004cvtree}, $d_2^*$ and $d_2^S$ \cite{wan2010alignment,reinert2009alignment}. However, we could not prove this claim theoretically.

In a series of papers, Kim and colleagues \cite{jun2010whole,sims2011whole,sims2009alignment,wu2009whole} investigated the optimal word length when using the Jensen--Shannon (JS) divergence between the word frequency vectors to measure the dissimilarity between two sequences. The lower limit of the word length was suggested as $\log_{L}(n)$, where $n$ is the average length of the sequences to be compared and $L$ is the alphabet size. To obtain an upper bound, they defined cumulative relative entropy (CRE) as follows. Let $F_k = (f_{\mathbf{w}}, \mathbf{w} \in \emph{A}^k)$ be the frequency vector of all the words of length $k$ and $\hat{F}_k  = (\hat{f}_{\mathbf{w}}, \mathbf{w} \in \mathcal{A}^k) $ be the corresponding expected frequency under the $k-2$-th order Markov chain.  \swallow{$\hat{f}_{\mathbf{w}} =\hat{f}_{\mathbf{w}}/(n-k+1)$.  For a given word $\mathbf{w} = w_1 w_2 \cdots w_k$, define $^{-}\mathbf{w} = w_2 \cdots w_k,  \mathbf{w}^{-} = w_1 w_2 \cdots w_{k-1}, ^{-}\mathbf{w}^{-} = w_2 w_3 \cdots w_{k-1}$. Then
\[ \hat{f}_{\mathbf{w}} = \frac{f_{^{-}\mathbf{w}} f_{\mathbf{w}^{-}}}{f_{^{-}\mathbf{w}^{-}}}. \]}
The CRE function is defined by
\[ \mathrm{CRE}(t) = \sum_{k = t}^{\infty} KL(\hat{F}_k, F_k), \]
where $KL$ is the Kullback-Leibler divergence. The upper bound of the optimal $k$ is the value of $t$ such that $\mathrm{CRE}(t)$ is close to zero. In practice, they used the $t$ such that $\mathrm{CRE}(t)$ is less than 10\% of the maximum CRE.  
For the pairwise comparison among a set of sequences, if the lengths of the sequences to be compared are not highly different,
the above approach will give similar lower and upper bounds for the optimal word length.
The final $k$-mer length can be chosen within the overlapping ranges of the optimal word length among the sequences.
If the sequences have highly different lengths, the authors suggested to divide the large genomes into blocks of equal length so that the sequences to be compared have similar length. They applied the method to investigate the relationships among the Escherichia coli/Shigella group \cite{sims2011whole}, prokaryotes \cite{jun2010whole}, and ds\text{DNA} viruses \cite{wu2009whole}.
Recently, Zhang et al. \cite{zhang2017viral} used the approach to investigate the relationship among close to 4,000 viruses with very different lengths. \swallow{They divided the virus genomes into four groups according to their lengths. First, based on the CRE criterion in \cite{sims2009alignment}, the upper limit for the optimal word length was determined to be between 9 to 13. Secondly, the authors defined the average common features (ACF) of a sequence as the average shared feature of this sequence with the other sequences. }

\swallow{
Another commonly used weighting method is to weight each $k$-mer by the TF-IDF (Frequency-Inverse Document Frequency) \cite{cong2017robust,cong2016exploring}. TF-IDF was first used for the classification of documents \cite{salton1986introduction,salton1971smart} and was recently used to study the relationship among molecular sequences. }

\section*{INTEGRATED SOFTWARE FOR ALIGNMENT-FREE SEQUENCE COMPARISON}

As reviewed in the above sections, a large number of alignment-free sequence comparison approaches have been developed and most of the individual studies have accompanying software tools available. To facilitate the use of the different alignment-free methods, a general-purpose alignment-free platform is desirable, which is expected to include the support of both assembled genome sequences and unassembled NGS shotgun reads as input, integration of exhaustive alignment-free sequence comparison measures, and visualization of results.

\begin{figure*}
\centering
\includegraphics[width=0.95\textwidth]{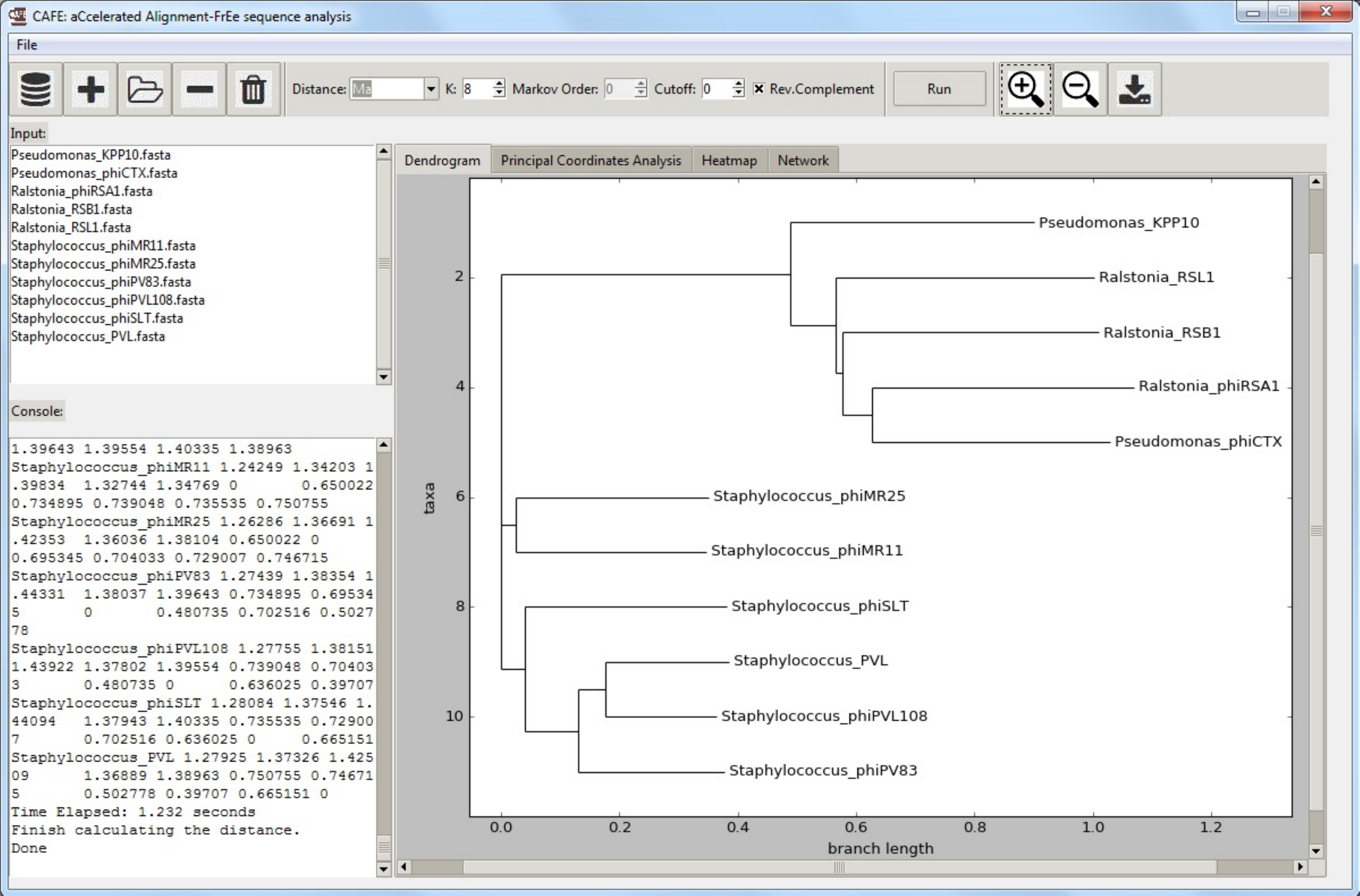}
\caption{Screenshot of the CAFE user interface based on a toy example comprising of $11$ bacterial genomes.
The user interface layout divides into six parts in terms of
functionality: (1) data selection toolbar (top left), (2)
dissimilarity setting toolbar (top middle), (3) image toolbar (top
right), (4) input data list (middle left), (5) run-time information
console (bottom left), and (6) visualized analyses (bottom right). }
\label{fig:cafe}
\end{figure*}

CAFE \cite{lu2017cafe} is a stand-alone alignment-free sequence comparison platform for studying the relationships among genomes and metagenomes through a user-friendly graphical user interface. Overall, CAFE integrates $28$ distinct alignment-free measures, including $10$ conventional measures based on $k$-mer counts (e.g., Euclidean, Manhattan, $d_{2}$, Jensen-Shannon divergence \cite{jun2010whole}, feature frequency profiles (FFP) \cite{sims2009alignment}, Co-phylog \cite{yi2013co}, etc.), $15$ measures based on presence/absence of $k$-mers (e.g., Jaccard, Hamming, etc.), and $3$ measures based on background adjusted $k$-mer counts (CVTree \cite{qi2004cvtree}, $d_{2}^{*}$ \cite{reinert2009alignment}, and $d_{2}^{S}$ \cite{reinert2009alignment}). All measures have been evaluated using whole primate and vertebrate genomes, whole microbial genomes, and NGS short reads from mammalian gut metagenomic samples. CAFE significantly speeds up the calculation of the background-adjusted measures such as CVTree, $d_{2}^{*}$, and $d_{2}^{S}$, with reduced memory requirements. Moreover, the resulting pairwise dissimilarities among the sequences form a symmetric distance matrix, which can be directly saved in a standard PHYLIP format (http://evolution.genetics.washington.edu/phylip/credits.html). CAFE also provides four types of built-in downstream visualized analyses, including clustering the sequences into dendrograms using the UPGMA algorithm, heatmap visualization of the matrix, projecting the matrix to a two-dimensional space using principal coordinate analysis (PCoA), and network display. A screenshot of CAFE is shown in \textbf{Figure ~\ref{fig:cafe}}.

Alternatively, Alfree \cite{zielezinski2017alignment} provides a publicly accessible web-based sequence comparison platform for studying the relationships among nucleotide and protein sequences. Alfree integrates $38$ popular alignment-free measures, including $25$ word-based measures (e.g., Euclidean, Minkowski, FFP, Jaccard, Hamming, etc.), $8$ Information-theoretic measures (e.g., Lempel-Ziv complexity \cite{otu2003new}, normalized compression distance \cite{li2004similarity}, etc.), $3$ graph-based measures \cite{yu2010novel}, and $2$ hybrid measures (i.e., Kullback--Leibler divergence \cite{wu2001statistical} and W-metric \cite{vinga2004comparative}). The majority of measures have been evaluated using simulated DNA sequences, primate mitochondrial genomes, prokaryotic genomes and proteomes, plant genomes, etc. Moreover, the resulting dissimilarities among the sequences are reported as phylogenetic trees, heat maps, and tables. 


With the advances of efficient and affordable sequencing technologies, the high volumes of sequence data have brought computational challenges even for alignment-free sequence comparison. This concern is alleviated by Mash \cite{ondov2016mash} that uses the MinHash dimensionality-reduction technique to reduce large amount of sequences to compressed sketch representations. Generally, Mash estimates the Jaccard distance between pairwise $k$-mer vectors in terms of compressed sketch representations, with moderate memory and computation overhead.
Similarly, kWIP \cite{murray2017kwip} counts $k$-mers, hashes them into a compressed sketch, and introduces an information-theoretic weighting to elevate the relevant $k$-mers against irrelevant ones. Finally, it computes the similarity as inner products of weighted frequency vectors, normalized by Shannon entropy.
In addition, Benoit et al. \cite{benoit2016multiple} developed a program, Simka, for fast calculation of various distance measures between sequences for $k$-mers up to 30 bps long.

\section*{DISCUSSION AND CONCLUSIONS}
With the development of NGS technologies, huge amounts of sequencing data can be generated efficiently and economically. Sequence comparison plays crucial roles to analyze the large amount of sequence data and to extract biological knowledge from them. Although alignment-based sequence comparison will continue to dominate molecular sequence analysis, alternative alignment-free sequence comparison has become increasingly important due to its efficiency in analyzing huge amount of sequence data as well as its comparable performance with alignment-based methods. In recent years, there is a surge of interest in using alignment-free sequence comparison approaches for investigating a variety of different problems including the study of evolutionary relationships of whole genome sequences and gene regulatory regions, comparison of metagenomes and metatranscriptomes, binning of contigs, detection of horizontal gene transfer, and virus-host infectious associations based on NGS data. Among the many types of alignment-free sequence comparison approaches, word-count based approaches are most popular due to their easy adaption to NGS data.

Most word-count based alignment-free approaches use the absolute word frequencies for sequence comparison. These approaches have the advantage of being simple, easy to calculate, and using less memory. On the other hand, relative word frequency based alignment-free methods that were originally developed by Karlin's group \cite{karlin1997compositional,karlin1995dinucleotide} and Hao's group \cite{qi2004cvtree,qi2004whole} and were recently revitalized by our group \cite{wan2010alignment,reinert2000probabilistic}  outperformed absolute word-count based approaches in all the applications we have investigated including the comparison of genomes \cite{song2013alignment,ren2016inference}, gene regulatory regions \cite{song2014new},  metagenomes \cite{jiang2012comparison}, and metatranscriptomics \cite{liao2016alignment}. They have also been used to improve the binning of contigs in metagenomes  \cite{wang2017improving} and to predict virus-host interactions \cite{ahlgren2017alignment}. By subtracting the expected word counts based on the background MC model from the observed word counts, the words distinguishing the sequences are strengthened while the weights of the irrelevant words are minimized resulting in the excellent performance of the background adjusted methods. However, the calculation of the background adjusted measures such as CVTree, $d_2^*$ and $d_2^S$ adds extra burdens in memory and computational speed. Further improvements to speed up the computation of these measures and to reduce memory are needed.

Although there have been some studies on the optimal choice of word length for some measures such as $\chi^2$-statistic \cite{bai2017optimal} and Jensen-Shannon entropy \cite{jun2010whole,sims2011whole,sims2009alignment,wu2009whole}, the optimal word length for many other measures is not known. In these studies, the optimal word length was determined by the individual sequences, not by the relationship among the sequences. We expect that for the comparison of highly divergent sequences, short word length should suffice, while for the study of closely related sequences, long word patterns are needed. However, no studies are available on the optimal word length considering the divergency among the sequences. A few recent studies \cite{murray2017kwip,ondov2016mash} used long words of length up to 30 bps and absolute word frequencies to compare genome sequences with excellent results and fast computation speed. It will be interesting to compare the performance of these approaches with the background adjusted measures with relatively short $k$-mers under realistic assumptions on sequencing errors and NGS data.

With the large number of alignment-free sequence comparison measures available, it is time to establish some benchmark data sets to evaluate the pros and cons of the different measures. Zielezinski et al. \cite{zielezinski2017alignment} built a benchmark data set of protein structures and evaluated a variety of different alignment-free sequence comparison measures and the Smith-Waterman algorithm.
Following up from their data set, there is a need for a collection of community-agreed data sets for the comparison of genomes, gene regulation regions, and metagenomes.

In summary, alignment-free sequence comparison methods have shown great promise for NGS data analysis as shown by many applications. They are generally computationally fast and use less memory compared to alignment based methods. Further studies on the choice of length of $k$-mers, differential weighting of the $k$-mers, and benchmark data sets are needed to explore the full potential of alignment-free methods.

\section*{DISCLOSURE STATEMENT}
The authors are not aware of any affiliations, memberships, funding, or financial holdings that might be perceived as affecting the objectivity of this review.

\section*{ACKNOWLEDGMENTS}
We thank Drs. Nathan A. Ahlgren, David Chew, Minghua Deng, Jed A. Fuhrman, Bai Jiang, Kai Song, Lin Wan, Michael S. Waterman, Xuegong Zhang, Ms. Weinan Liao and Ms. Kun Wang for collaborations on the investigation of alignment-free sequence comparison and applications.   The preparation of the manuscript was supported by US NSF National Science Foundation (NSF) [DMS-1518001] and National Institutes of Health [R01GM120624]. Dr. Ying Wang was supported by National Natural Science Foundation of
China (61673324, 61503314),  China Scholarship Council (201606315011)
and Natural Science Foundation of Fujian (2016 J01316).

\bibliography{alignment_free_review_revised_arXiv}

\begin{thebibliography}{160}
\expandafter\ifx\csname natexlab\endcsname\relax\def\natexlab#1{#1}\fi

\bibitem{smith1981identification}
Smith TF, Waterman MS. 1981.
Identification of common molecular subsequences.
\textit{Journal of Molecular Biology} 147:195--197

\bibitem{altschul1990basic}
Altschul SF, Gish W, Miller W, Myers EW, Lipman DJ. 1990.
Basic local alignment search tool.
\textit{Journal of Molecular Biology} 215:403--410

\bibitem{kent2002blat}
Kent WJ. 2002.
{BLAT}: the {BLAST}-like alignment tool.
\textit{Genome Research} 12:656--664

\bibitem{wang2009fungal}
Wang H, Xu Z, Gao L, Hao B. 2009.
{A fungal phylogeny based on 82 complete genomes using the composition vector
  method}.
\textit{BMC Evolutionary Biology} 9:195

\bibitem{jun2010whole}
Jun S, Sims G, Wu G, Kim S. 2010.
{Whole-proteome phylogeny of prokaryotes by feature frequency profiles: An
  alignment-free method with optimal feature resolution}.
\textit{Proceedings of the National Academy of Sciences of the United States of
  America} 107:133--138

\bibitem{sims2011whole}
Sims GE, Kim SH. 2011.
Whole-genome phylogeny of escherichia coli/shigella group by feature frequency
  profiles (ffps).
\textit{Proceedings of the National Academy of Sciences} 108:8329--8334

\bibitem{blaisdell1986measure}
Blaisdell B. 1986.
{A measure of the similarity of sets of sequences not requiring sequence
  alignment}.
\textit{Proceedings of the National Academy of Sciences of the United States of
  America} 83:5155--5159

\bibitem{blaisdell1985markov}
Blaisdell BE. 1985.
\mbox{M}arkov chain analysis finds a significant influence of neighboring bases
  on the occurrence of a base in eucaryotic nuclear \mbox{DNA} sequences both
  protein-coding and noncoding.
\textit{Journal of Molecular Evolution} 21:278--288

\bibitem{torney1990computation}
Torney D, Burks C, Davison D, Sirotkin K. 1990.
{Computation of d2: A measure of sequence dissimilarity}.
\textit{Computers and \mbox{DNA}} :109--125

\bibitem{wan2010alignment}
Wan L, Reinert G, Sun F, Waterman M. 2010.
Alignment-free sequence comparison (\mbox{II}): Theoretical power of comparison
  statistics.
\textit{Journal of Computational Biology} 17:1467--1490

\bibitem{reinert2009alignment}
Reinert G, Chew D, Sun FZ, Waterman MS. 2009.
Alignment-free sequence comparison ({I}): Statistics and power.
\textit{Journal of Computational Biology} 16:1615--1634

\bibitem{sims2009alignment}
Sims G, Jun S, Wu G, Kim S. 2009.
{Alignment-free genome comparison with feature frequency profiles ({FFP}) and
  optimal resolutions}.
\textit{Proceedings of the National Academy of Sciences of the United States of
  America} 106:2677--2682

\bibitem{qi2004cvtree}
Qi J, Luo H, Hao B. 2004.
{CVTree: a phylogenetic tree reconstruction tool based on whole genomes}.
\textit{Nucleic Acids Research} 32:W45

\bibitem{ulitsky2006average}
Ulitsky I, Burstein D, Tuller T, Chor B. 2006.
{The average common substring approach to phylogenomic reconstruction}.
\textit{Journal of Computational Biology} 13:336--350

\bibitem{yang2016estimator}
Yang L, Zhang X, Fu H, Yang C. 2016.
An estimator for local analysis of genome based on the minimal absent word.
\textit{Journal of Theoretical Biology} 395:23--30

\bibitem{yang2012alignment}
Yang L, Zhang X, Zhu H. 2012.
Alignment free comparison: Similarity distribution between the dna primary
  sequences based on the shortest absent word.
\textit{Journal of Theoretical Biology} 295:125--131

\bibitem{yang2013large}
Yang L, Zhang X, Wang T, Zhu H. 2013.
Large local analysis of the unaligned genome and its application.
\textit{Journal of Computational Biology} 20:19--29

\bibitem{almeida2001analysis}
Almeida JS, Carrico JA, Maretzek A, Noble PA, Fletcher M. 2001.
Analysis of genomic sequences by chaos game representation.
\textit{Bioinformatics} 17:429--437

\bibitem{wang2005spectrum}
Wang Y, Hill K, Singh S, Kari L. 2005.
The spectrum of genomic signatures: from dinucleotides to chaos game
  representation.
\textit{Gene} 346:173--185

\bibitem{jeffrey1990chaos}
Jeffrey HJ. 1990.
Chaos game representation of gene structure.
\textit{Nucleic Acids Research} 18:2163--2170

\bibitem{yau2008protein}
Yau SST, Yu C, He R. 2008.
A protein map and its application.
\textit{DNA and Cell Biology} 27:241--250

\bibitem{yin2015improved}
Yin C, Yau SST. 2015.
An improved model for whole genome phylogenetic analysis by fourier transform.
\textit{Journal of Theoretical Biology} 382:99--110

\bibitem{vinga2013information}
Vinga S. 2013.
Information theory applications for biological sequence analysis.
\textit{Briefings in Bioinformatics} 15:376--389

\bibitem{almeida2013sequence}
Almeida JS. 2013.
Sequence analysis by iterated maps, a review.
\textit{Briefings in Bioinformatics} 15:369--375

\bibitem{zielezinski2017alignment}
Zielezinski A, Vinga S, Almeida J, Karlowski WM. 2017.
Alignment-free sequence comparison: benefits, applications, and tools.
\textit{Genome Biology} 18:186

\bibitem{bonham2013alignment}
Bonham-Carter O, Steele J, Bastola D. 2013.
Alignment-free genetic sequence comparisons: a review of recent approaches by
  word analysis.
\textit{Briefings in Bioinformatics} 15:890--905

\bibitem{song2014new}
Song K, Ren J, Reinert G, Deng M, Waterman MS, Sun F. 2014.
New developments of alignment-free sequence comparison: measures, statistics
  and next-generation sequencing.
\textit{Briefings in Bioinformatics} 15:343--353

\bibitem{vinga2003alignment}
Vinga S, Almeida J. 2003.
{Alignment-free sequence comparison--a review}.
\textit{Bioinformatics} 19:513--523

\bibitem{li2010composition}
Li Q, Xu Z, Hao B. 2010.
Composition vector approach to whole-genome-based prokaryotic phylogeny:
  success and foundations.
\textit{Journal of Biotechnology} 149:115--119

\bibitem{marccais2011fast}
Mar{\c{c}}ais G, Kingsford C. 2011.
A fast, lock-free approach for efficient parallel counting of occurrences of
  k-mers.
\textit{Bioinformatics} 27:764--770

\bibitem{rizk2013dsk}
Rizk G, Lavenier D, Chikhi R. 2013.
Dsk: k-mer counting with very low memory usage.
\textit{Bioinformatics} 29:652--653

\bibitem{deorowicz2015kmc}
Deorowicz S, Kokot M, Grabowski S, Debudaj-Grabysz A. 2015.
Kmc 2: fast and resource-frugal k-mer counting.
\textit{Bioinformatics} 31:1569--1576

\bibitem{sobieski198416s}
Sobieski JM, Nan~Chen K, Filiatreau JC, Pickett MH, Fox GE. 1984.
16s rrna oligonucleotide catalog data base.
\textit{Nucleic Acids Res.} 12:141--148

\bibitem{fox1980phylogeny}
Fox Gca, Stackebrandt E, Hespell R, Gibson J, Maniloff J, et~al. 1980.
The phylogeny of prokaryotes.
\textit{Science} 209:457--463

\bibitem{fox1977classification}
Fox GE, Magrum LJ, Balch WE, Wolfe RS, Woese CR. 1977.
Classification of methanogenic bacteria by 16s ribosomal rna characterization.
\textit{Proceedings of the National Academy of Sciences} 74:4537--4541

\bibitem{woese1985phylogenetic}
Woese C, Stackebrandt E, Macke T, Fox G. 1985.
A phylogenetic definition of the major eubacterial taxa.
\textit{Systematic and Applied Microbiology} 6:143--151

\bibitem{mcgill1986characteristic}
McGill TJ, Jurka J, Sobieski JM, Pickett MH, Woese CR, Fox GE. 1986.
Characteristic archaebacterial 16s rrna oligonucleotides.
\textit{Systematic and Applied Microbiology} 7:194--197

\bibitem{woese1985mycoplasmas}
Woese C, Stackebrandt E, Ludwig W. 1985.
What are mycoplasmas: the relationship of tempo and mode in bacterial
  evolution.
\textit{Journal of Molecular Evolution} 21:305--316

\bibitem{fox1977comparative}
FOX GE, Pechman KR, Woese CR. 1977.
Comparative cataloging of 16s ribosomal ribonucleic acid: molecular approach to
  procaryotic systematics.
\textit{International Journal of Systematic and Evolutionary Microbiology}
  27:44--57

\bibitem{woese1987bacterial}
Woese CR. 1987.
Bacterial evolution.
\textit{Microbiological Reviews} 51:221

\bibitem{ragan2014molecular}
Ragan MA, Bernard G, Chan CX. 2014.
Molecular phylogenetics before sequences: oligonucleotide catalogs as k-mer
  spectra.
\textit{RNA Biology} 11:176--185

\bibitem{karlin1997compositional}
Karlin S, Mr{\'a}zek J. 1997.
{Compositional differences within and between eukaryotic genomes}.
\textit{Proceedings of the National Academy of Sciences of the United States of
  America} 94:10227--10232

\bibitem{karlin1995dinucleotide}
Karlin S, Burge C. 1995.
{Dinucleotide relative abundance extremes: a genomic signature}.
\textit{Trends in Genetics} 11:283--290

\bibitem{bernard2016alignment}
Bernard G, Chan CX, Ragan MA. 2016.
Alignment-free microbial phylogenomics under scenarios of sequence divergence,
  genome rearrangement and lateral genetic transfer.
\textit{Scientific Reports} 6:28970

\bibitem{chan2014inferring}
Chan CX, Bernard G, Poirion O, Hogan JM, Ragan MA. 2014.
Inferring phylogenies of evolving sequences without multiple sequence
  alignment.
\textit{Scientific Reports} 4

\bibitem{lu2017cafe}
Lu YY, Tang K, Ren J, Fuhrman JA, Waterman MS, Sun F. 2017.
Cafe: accelerated alignment-free sequence analysis.
\textit{Nucleic Acids Research} :10.1093/nar/gkx351

\bibitem{narlikar2013one}
Narlikar L, Mehta N, Galande S, Arjunwadkar M. 2013.
One size does not fit all: On how markov model order dictates performance of
  genomic sequence analyses.
\textit{Nucleic Acids Research} 41:1416--1424

\bibitem{liu2011new}
Liu X, Wan L, Li J, Reinert G, Waterman M, Sun F. 2011.
New powerful statistics for alignment-free sequence comparison under a pattern
  transfer model.
\textit{Journal of Theoretical Biology} 284:106--116

\bibitem{song2013alignment}
Song K, Ren J, Zhai Z, Liu X, Deng M, Sun F. 2013.
Alignment-free sequence comparison based on next-generation sequencing reads.
\textit{Journal of Computational Biology} 20:64--79

\bibitem{ren2013multiple}
Ren J, Song K, Sun F, Deng M, Reinert G. 2013.
Multiple alignment-free sequence comparison.
\textit{Bioinformatics} 29:2690--2698

\bibitem{qi2004whole}
Qi J, Wang B, Hao BI. 2004.
Whole proteome prokaryote phylogeny without sequence alignment: a k-string
  composition approach.
\textit{Journal of Molecular Evolution} 58:1--11

\bibitem{teeling2004tetra}
Teeling H, Waldmann J, Lombardot T, Bauer M, Gl{\"o}ckner FO. 2004.
Tetra: a web-service and a stand-alone program for the analysis and comparison
  of tetranucleotide usage patterns in dna sequences.
\textit{BMC Bioinformatics} 5:163

\bibitem{pride2006evidence}
Pride DT, Wassenaar TM, Ghose C, Blaser MJ. 2006.
Evidence of host-virus co-evolution in tetranucleotide usage patterns of
  bacteriophages and eukaryotic viruses.
\textit{BMC Genomics} 7:8

\bibitem{willner2009metagenomic}
Willner D, Thurber RV, Rohwer F. 2009.
Metagenomic signatures of 86 microbial and viral metagenomes.
\textit{Environmental Microbiology} 11:1752--1766

\bibitem{almagor1983markov}
Almagor H. 1983.
A \mbox{M}arkov analysis of \mbox{DNA} sequences.
\textit{Journal of Theoretical Biology} 104:633--645

\bibitem{pevzner1989linguistics}
Pevzner PA, Borodovsky MY, Mironov AA. 1989.
Linguistics of nucleotide sequences i: the significance of deviations from mean
  statistical characteristics and prediction of the frequencies of occurrence
  of words.
\textit{Journal of Biomolecular Structure and Dynamics} 6:1013--1026

\bibitem{hong1990prediction}
Hong J. 1990.
Prediction of oligonucleotide frequencies based upon dinucleotide frequencies
  obtained from the nearest neighbor analysis.
\textit{Nucleic Acids Research} 18:1625--1628

\bibitem{arnold1988mono}
Arnold J, Cuticchia AJ, Newsome DA, Jennings WW, Ivarie R. 1988.
Mono-through hexanucleotide composition of the sense strand of yeast
  \mbox{DNA}: a \mbox{M}arkov chain analysis.
\textit{Nucleic Acids Research} 16:7145--7158

\bibitem{avery1987analysis}
Avery PJ. 1987.
The analysis of intron data and their use in the detection of short signals.
\textit{Journal of Molecular Evolution} 26:335--340

\bibitem{hoel1954test}
Hoel PG. 1954.
A test for \mbox{M}arkov chains.
\textit{Biometrika} 41:430--433

\bibitem{anderson1957statistical}
Anderson TW, Goodman LA. 1957.
Statistical inference about \mbox{M}arkov chains.
\textit{The Annals of Mathematical Statistics} 28:89--110

\bibitem{avery1999fitting}
Avery PJ, Henderson DA. 1999.
Fitting \mbox{M}arkov chain models to discrete state series such as \mbox{DNA}
  sequences.
\textit{Journal of the Royal Statistical Society: Series C (Applied
  Statistics)} 48:53--61

\bibitem{billingsley1961statisticalb}
Billingsley P. 1961{\natexlab{a}}.
\textit{Statistical Inference for \mbox{M}arkov Processes}, vol.~2.
University of Chicago Press Chicago

\bibitem{billingsley1961statisticala}
Billingsley P. 1961{\natexlab{b}}.
Statistical methods in \mbox{M}arkov chains.
\textit{The Annals of Mathematical Statistics} 32:12--40

\bibitem{waterman1995introduction}
Waterman MS. 1995.
\textit{Introduction to Computational Biology: Maps, Sequences and Genomes}.
Chapman \& Hall/CRC Interdisciplinary Statistics. Taylor \& Francis

\bibitem{reinert2000probabilistic}
Reinert G, Schbath S, Waterman M. 2000.
Probabilistic and statistical properties of words: an overview.
\textit{Journal of Computational Biology} 7:1--46

\bibitem{reinert2005statistics}
Reinert G, Schbath S, Waterman MS. 2005.
Statistics on words with applications to biological sequences.
\textit{Lothaire: Applied Combinatorics on Words, J. Berstel and D. Perrin,
  eds.} 105:251--328

\bibitem{ewens2005statistical}
Ewens WJ, Grant GR. 2005.
\textit{Statistical methods in bioinformatics: an introduction}.
Springer

\bibitem{menendez2011testing}
Men{\'e}ndez ML, Pardo L, Pardo M, Zografos K. 2011.
Testing the order of \mbox{M}arkov dependence in \mbox{DNA} sequences.
\textit{Methodology and Computing in Applied Probability} 13:59--74

\bibitem{papapetrou2013markov}
Papapetrou M, Kugiumtzis D. 2013.
Markov chain order estimation with conditional mutual information.
\textit{Physica A: Statistical Mechanics and its Applications} 392:1593--1601

\bibitem{morvai2005order}
Morvai G, Weiss B. 2005.
Order estimation of \mbox{M}arkov chains.
\textit{Information Theory, IEEE Transactions on} 51:1496--1497

\bibitem{peres2005two}
Peres Y, Shields P. 2005.
Two new \mbox{M}arkov order estimators.
\textit{arXiv preprint math/0506080}

\bibitem{dalevi2006new}
Dalevi D, Dubhashi D, Hermansson M. 2006.
A new order estimator for fixed and variable length \mbox{M}arkov models with
  applications to \mbox{DNA} sequence similarity.
\textit{Statistical Applications in Genetics and Molecular Biology} 5

\bibitem{baigorri2009markov}
Baigorri A, Gon{\c{c}}alves C, Resende P. 2009.
Markov chain order estimation and $\chi$ 2- divergence measure.
\textit{arXiv preprint arXiv:0910.0264}

\bibitem{besag2013exact}
Besag J, Mondal D. 2013.
Exact goodness-of-fit tests for \mbox{M}arkov chains.
\textit{Biometrics} 69:488--496

\bibitem{tong1975determination}
Tong H. 1975.
Determination of the order of a \mbox{M}arkov chain by \mbox{A}kaike's
  information criterion.
\textit{Journal of Applied Probability} 12:488--497

\bibitem{hurvich1995model}
Hurvich CM, Tsai CL. 1995.
Model selection for extended quasi-likelihood models in small samples.
\textit{Biometrics} :1077--1084

\bibitem{zhao2001determination}
Zhao LC, Dorea CCY, Gon{\c{c}}alves CR. 2001.
On determination of the order of a \mbox{M}arkov chain.
\textit{Statistical Inference for Stochastic Processes} 4:273--282

\bibitem{dorea2006convergence}
Dorea C, Lopes J. 2006.
Convergence rates for \mbox{M}arkov chain order estimates using edc criterion.
\textit{Bulletin of the Brazilian Mathematical Society} 37:561--570

\bibitem{katz1981some}
Katz RW. 1981.
On some criteria for estimating the order of a \mbox{M}arkov chain.
\textit{Technometrics} 23:243--249

\bibitem{ren2016inference}
Ren J, Song K, Deng M, Reinert G, Cannon CH, Sun F. 2016.
Inference of markovian properties of molecular sequences from ngs data and
  applications to comparative genomics.
\textit{Bioinformatics} 32:993--1000

\bibitem{lander1988genomic}
Lander ES, Waterman MS. 1988.
Genomic mapping by fingerprinting random clones: a mathematical analysis.
\textit{Genomics} 2:231--239

\bibitem{burden2012alignment}
Burden CJ, Jing J, Wilson SR. 2012.
Alignment-free sequence comparison for biologically realistic sequences of
  moderate length.
\textit{Statistical Applications in Genetics and Molecular Biology} 11:1--28

\bibitem{cannon2010assembly}
Cannon CH, Kua CS, Zhang D, Harting J. 2010.
{Assembly free comparative genomics of short-read sequence data discovers the
  needles in the haystack}.
\textit{Molecular Ecology} 19:146--160

\bibitem{miller200728}
Miller W, Rosenbloom K, Hardison R, Hou M, Taylor J, et~al. 2007.
28-way vertebrate alignment and conservation track in the \mbox{UCSC} genome
  browser.
\textit{Genome Research} 17:1797--1808

\bibitem{bernard2016recapitulating}
Bernard G, Ragan MA, Chan CX. 2016.
Recapitulating phylogenies using k-mers: from trees to networks.
\textit{F1000Research} 5

\bibitem{yi2013co}
Yi H, Jin L. 2013.
Co-phylog: an assembly-free phylogenomic approach for closely related
  organisms.
\textit{Nucleic Acids Research} 41:e75

\bibitem{leimeister2014fast}
Leimeister CA, Boden M, Horwege S, Lindner S, Morgenstern B. 2014.
Fast alignment-free sequence comparison using spaced-word frequencies.
\textit{Bioinformatics} 30:1991--1999

\bibitem{fan2015assembly}
Fan H, Ives AR, Surget-Groba Y, Cannon CH. 2015.
An assembly and alignment-free method of phylogeny reconstruction from
  next-generation sequencing data.
\textit{BMC Genomics} 16:522

\bibitem{cattaneo2017effective}
Cattaneo G, Petrillo UF, Giancarlo R, Roscigno G. 2017.
An effective extension of the applicability of alignment-free biological
  sequence comparison algorithms with hadoop.
\textit{The Journal of Supercomputing} 73:1467--1483

\bibitem{bernard2017alignment}
Bernard G, Chan CX, Chan Yb, Chua XY, Cong Y, et~al. 2017.
Alignment-free inference of hierarchical and reticulate phylogenomic
  relationships.
\textit{Briefings in Bioinformatics}

\bibitem{rappe2003uncultured}
Rapp{\'e} MS, Giovannoni SJ. 2003.
The uncultured microbial majority.
\textit{Annual Reviews in Microbiology} 57:369--394

\bibitem{dutilh2014highly}
Dutilh BE, Cassman N, McNair K, Sanchez SE, Silva GG, et~al. 2014.
A highly abundant bacteriophage discovered in the unknown sequences of human
  faecal metagenomes.
\textit{Nature Communications} 5

\bibitem{norman2015disease}
Norman JM, Handley SA, Baldridge MT, Droit L, Liu CY, et~al. 2015.
{Disease-specific alterations in the enteric virome in inflammatory bowel
  disease}.
\textit{Cell} 160:447--460

\bibitem{reyes2015gut}
Reyes A, Blanton LV, Cao S, Zhao G, Manary M, et~al. 2015.
{Gut DNA viromes of Malawian twins discordant for severe acute malnutrition}.
\textit{Proceedings of the National Academy of Sciences} 112:11941--11946

\bibitem{minot2011human}
Minot S, Sinha R, Chen J, Li H, Keilbaugh SA, et~al. 2011.
{The human gut virome: inter-individual variation and dynamic response to
  diet}.
\textit{Genome Research} 21:1616--1625

\bibitem{waller2014classification}
Waller AS, Yamada T, Kristensen DM, Kultima JR, Sunagawa S, et~al. 2014.
{Classification and quantification of bacteriophage taxa in human gut
  metagenomes}.
\textit{The ISME Journal} 8:1391--1402

\bibitem{brum2015patterns}
Brum JR, Ignacio-Espinoza JC, Roux S, Doulcier G, Acinas SG, et~al. 2015.
Patterns and ecological drivers of ocean viral communities.
\textit{Science} 348:1261498

\bibitem{reyes2010viruses}
Reyes A, Haynes M, Hanson N, Angly FE, Heath AC, et~al. 2010.
Viruses in the faecal microbiota of monozygotic twins and their mothers.
\textit{Nature} 466:334--338

\bibitem{zhang2005rna}
Zhang T, Breitbart M, Lee WH, Run JQ, Wei CL, et~al. 2005.
Rna viral community in human feces: prevalence of plant pathogenic viruses.
\textit{PLoS Biol} 4:e3

\bibitem{pearce2012metagenomic}
Pearce DA, Newsham KK, Thorne MA, Calvo-Bado L, Krsek M, et~al. 2012.
Metagenomic analysis of a southern maritime antarctic soil.
\textit{Front. Microbiol.} 3:403. 10.3389/fmicb.2012.00403

\bibitem{adriaenssens2015metagenomic}
Adriaenssens EM, Van~Zyl L, De~Maayer P, Rubagotti E, Rybicki E, et~al. 2015.
Metagenomic analysis of the viral community in namib desert hypoliths.
\textit{Environmental Microbiology} 17:480--495

\bibitem{zablocki2014high}
Zablocki O, van Zyl L, Adriaenssens EM, Rubagotti E, Tuffin M, et~al. 2014.
High-level diversity of tailed phages, eukaryote-associated viruses, and
  virophage-like elements in the metaviromes of antarctic soils.
\textit{Applied and Environmental Microbiology} 80:6888--6897

\bibitem{roux2015virsorter}
Roux S, Enault F, Hurwitz BL, Sullivan MB. 2015{\natexlab{a}}.
Virsorter: mining viral signal from microbial genomic data.
\textit{PeerJ} 3:e985

\bibitem{edwards2016computational}
Edwards RA, McNair K, Faust K, Raes J, Dutilh BE. 2016.
Computational approaches to predict bacteriophage--host relationships.
\textit{FEMS Microbiology Reviews} 40:258--272

\bibitem{roux2015viral}
Roux S, Hallam SJ, Woyke T, Sullivan MB. 2015{\natexlab{b}}.
Viral dark matter and virus--host interactions resolved from publicly available
  microbial genomes.
\textit{Elife} 4:e08490

\bibitem{ahlgren2017alignment}
Ahlgren NA, Ren J, Lu YY, Fuhrman JA, Sun F. 2017.
Alignment-free $ d_2^* $ oligonucleotide frequency dissimilarity measure
  improves prediction of hosts from metagenomically-derived viral sequences.
\textit{Nucleic Acids Research} 45:39--53

\bibitem{galiez2017wish}
Galiez C, Siebert M, Enault F, Vincent J, S{\"o}ding J. 2017.
Wish: who is the host? predicting prokaryotic hosts from metagenomic phage
  contigs.
\textit{Bioinformatics} 33:3113--3114

\bibitem{paez2016uncovering}
Paez-Espino D, Eloe-Fadrosh EA, Pavlopoulos GA, Thomas AD, Huntemann M, et~al.
  2016.
Uncovering earth's virome.
\textit{Nature} 536

\bibitem{lima2008reticulate}
Lima-Mendez G, Van~Helden J, Toussaint A, Leplae R. 2008.
Reticulate representation of evolutionary and functional relationships between
  phage genomes.
\textit{Molecular Biology and Evolution} 25:762--777

\bibitem{shapiro2017gene}
Shapiro JW, Putonti C. 2017.
Gene networks provide a high-resolution view of bacteriophage ecology.
\textit{bioRxiv} :148668

\bibitem{villarroel2016hostphinder}
Villarroel J, Kleinheinz KA, Jurtz VI, Zschach H, Lund O, et~al. 2016.
Hostphinder: a phage host prediction tool.
\textit{Viruses} 8:116

\bibitem{zhang2017prediction}
Zhang M, Yang L, Ren J, Ahlgren NA, Fuhrman JA, Sun F. 2017{\natexlab{a}}.
Prediction of virus-host infectious association by supervised learning methods.
\textit{BMC Bioinformatics} 18:60

\bibitem{liao2016alignment}
Liao W, Ren J, Wang K, Wang S, Zeng F, et~al. 2016.
Alignment-free transcriptomic and metatranscriptomic comparison using
  sequencing signatures with variable length markov chains.
\textit{Scientific Reports} 6:37243

\bibitem{buhlmann1999variable}
B{\"u}hlmann P, Wyner AJ, et~al. 1999.
Variable length markov chains.
\textit{The Annals of Statistics} 27:480--513

\bibitem{rissanen1983universal}
Rissanen J. 1983.
A universal data compression system.
\textit{IEEE Transactions on information theory} 29:656--664

\bibitem{kullback1951information}
Kullback S, Leibler RA. 1951.
On information and sufficiency.
\textit{The annals of mathematical statistics} 22:79--86

\bibitem{akaike1987factor}
Akaike H. 1987.
Factor analysis and aic.
\textit{Psychometrika} 52:317--332

\bibitem{jiang2012comparison}
Jiang B, Song K, Ren J, Deng M, Sun F, Zhang X. 2012.
Comparison of metagenomic samples using sequence signatures.
\textit{BMC Genomics} 13:730

\bibitem{behnam2014amordad}
Behnam E, Smith AD. 2014.
The amordad database engine for metagenomics.
\textit{Bioinformatics} 30:2949--2955

\bibitem{benoit2016multiple}
Benoit G, Peterlongo P, Mariadassou M, Drezen E, Schbath S, et~al. 2016.
Multiple comparative metagenomics using multiset k-mer counting.
\textit{PeerJ Computer Science} 2:e94

\bibitem{wang2017improving}
Wang Y, Wang K, Lu YY, Sun F. 2017.
Improving contig binning of metagenomic data using d2s oligonucleotide
  frequency dissimilarity.
\textit{BMC Bioinformatics} 18:425

\bibitem{mande2012classification}
Mande SS, Mohammed MH, Ghosh TS. 2012.
Classification of metagenomic sequences: methods and challenges.
\textit{Briefings in bioinformatics} 13:669--681

\bibitem{dick2009community}
Dick GJ, Andersson AF, Baker BJ, Simmons SL, Thomas BC, et~al. 2009.
Community-wide analysis of microbial genome sequence signatures.
\textit{Genome biology} 10:R85

\bibitem{leung2011robust}
Leung HC, Yiu SM, Yang B, Peng Y, Wang Y, et~al. 2011.
A robust and accurate binning algorithm for metagenomic sequences with
  arbitrary species abundance ratio.
\textit{Bioinformatics} 27:1489--1495

\bibitem{kislyuk2009unsupervised}
Kislyuk A, Bhatnagar S, Dushoff J, Weitz JS. 2009.
Unsupervised statistical clustering of environmental shotgun sequences.
\textit{BMC bioinformatics} 10:316

\bibitem{wu2011novel}
Wu YW, Ye Y. 2011.
A novel abundance-based algorithm for binning metagenomic sequences using
  l-tuples.
\textit{Journal of Computational Biology} 18:523--534

\bibitem{wang2015mbbc}
Wang Y, Hu H, Li X. 2015.
Mbbc: an efficient approach for metagenomic binning based on clustering.
\textit{BMC bioinformatics} 16:36

\bibitem{wu2014maxbin}
Wu YW, Tang YH, Tringe SG, Simmons BA, Singer SW. 2014.
Maxbin: an automated binning method to recover individual genomes from
  metagenomes using an expectation-maximization algorithm.
\textit{Microbiome} 2:26

\bibitem{lin2016accurate}
Lin HH, Liao YC. 2016.
Accurate binning of metagenomic contigs via automated clustering sequences
  using information of genomic signatures and marker genes.
\textit{Scientific reports} 6

\bibitem{strous2012binning}
Strous M, Kraft B, Bisdorf R, Tegetmeyer HE. 2012.
The binning of metagenomic contigs for microbial physiology of mixed cultures.
\textit{Frontiers in microbiology} 3

\bibitem{kelley2010clustering}
Kelley DR, Salzberg SL. 2010.
Clustering metagenomic sequences with interpolated markov models.
\textit{BMC bioinformatics} 11:544

\bibitem{pal2005adaptive}
P{\'a}l C, Papp B, Lercher MJ. 2005.
Adaptive evolution of bacterial metabolic networks by horizontal gene transfer.
\textit{Nature Genetics} 37:1372

\bibitem{gyles2014horizontally}
Gyles C, Boerlin P. 2014.
Horizontally transferred genetic elements and their role in pathogenesis of
  bacterial disease.
\textit{Veterinary Pathology} 51:328--340

\bibitem{ravenhall2015inferring}
Ravenhall M, {\v{S}}kunca N, Lassalle F, Dessimoz C. 2015.
Inferring horizontal gene transfer.
\textit{PLoS Computational Biology} 11:e1004095

\bibitem{lu2016computational}
Lu B, Leong HW. 2016.
Computational methods for predicting genomic islands in microbial genomes.
\textit{Computational and Structural Biotechnology Journal} 14:200--206

\bibitem{keeling2008horizontal}
Keeling PJ, Palmer JD. 2008.
Horizontal gene transfer in eukaryotic evolution.
\textit{Nature Reviews Genetics} 9:605

\bibitem{cong2017robust}
Cong Y, Chan Yb, Phillips CA, Langston MA, Ragan MA. 2017.
Robust inference of genetic exchange communities from microbial genomes using
  tf-idf.
\textit{Frontiers in Microbiology} 8

\bibitem{cong2016novel}
Cong Y, Chan Yb, Ragan MA. 2016{\natexlab{a}}.
A novel alignment-free method for detection of lateral genetic transfer based
  on {TF-IDF}.
\textit{Scientific Reports} 6

\bibitem{cong2016exploring}
Cong Y, Chan Yb, Ragan MA. 2016{\natexlab{b}}.
Exploring lateral genetic transfer among microbial genomes using tf-idf.
\textit{Scientific Reports} 6:29319

\bibitem{dufraigne2005detection}
Dufraigne C, Fertil B, Lespinats S, Giron A, Deschavanne P. 2005.
Detection and characterization of horizontal transfers in prokaryotes using
  genomic signature.
\textit{Nucleic Acids Research} 33:e6--e6

\bibitem{rajan2007identification}
Rajan I, Aravamuthan S, Mande SS. 2007.
Identification of compositionally distinct regions in genomes using the
  centroid method.
\textit{Bioinformatics} 23:2672--2677

\bibitem{tsirigos2005new}
Tsirigos A, Rigoutsos I. 2005.
A new computational method for the detection of horizontal gene transfer
  events.
\textit{Nucleic Acids Research} 33:922--933

\bibitem{becq2010benchmark}
Becq J, Churlaud C, Deschavanne P. 2010.
A benchmark of parametric methods for horizontal transfers detection.
\textit{PLoS One} 5:e9989

\bibitem{karlin2001detecting}
Karlin S. 2001.
Detecting anomalous gene clusters and pathogenicity islands in diverse
  bacterial genomes.
\textit{Trends in Microbiology} 9:335--343

\bibitem{tamames2008estimating}
Tamames J, Moya A. 2008.
Estimating the extent of horizontal gene transfer in metagenomic sequences.
\textit{BMC Genomics} 9:136

\bibitem{goke2012estimation}
G{\"o}ke J, Schulz MH, Lasserre J, Vingron M. 2012.
Estimation of pairwise sequence similarity of mammalian enhancers with word
  neighbourhood counts.
\textit{Bioinformatics} 28:656--663

\bibitem{horwege2014spaced}
Horwege S, Lindner S, Boden M, Hatje K, Kollmar M, et~al. 2014.
Spaced words and kmacs: fast alignment-free sequence comparison based on
  inexact word matches.
\textit{Nucleic Acids Research} 42:W7--W11

\bibitem{patil2013alignment}
Patil KR, McHardy AC. 2013.
Alignment-free genome tree inference by learning group-specific distance
  metrics.
\textit{Genome Biology and Evolution} 5:1470--1484

\bibitem{qian2017weighted}
Qian K, Luan Y. 2017.
Weighted measures based on maximizing deviation for alignment-free sequence
  comparison.
\textit{Physica A: Statistical Mechanics and its Applications} 481:235--242

\bibitem{murray2017kwip}
Murray KD, Webers C, Ong CS, Borevitz J, Warthmann N. 2017.
kwip: The k-mer weighted inner product, a de novo estimator of genetic
  similarity.
\textit{PLOS Computational Biology} 13:e1005727

\bibitem{ondov2016mash}
Ondov BD, Treangen TJ, Melsted P, Mallonee AB, Bergman NH, et~al. 2016.
Mash: fast genome and metagenome distance estimation using minhash.
\textit{Genome Biology} 17:132

\bibitem{bai2017optimal}
Bai X, Tang K, Ren J, Waterman M, Sun F. 2017.
Optimal choice of word length when comparing two markov sequences using a
  $\chi$ 2-statistic.
\textit{BMC Genomics} 18:732

\bibitem{wu2009whole}
Wu G, Jun S, Sims G, Kim S. 2009.
{Whole-proteome phylogeny of large ds\mbox{DNA} virus families by an
  alignment-free method}.
\textit{Proceedings of the National Academy of Sciences of the United States of
  America} 106:12826--12831

\bibitem{zhang2017viral}
Zhang Q, Jun SR, Leuze M, Ussery D, Nookaew I. 2017{\natexlab{b}}.
Viral phylogenomics using an alignment-free method: A three-step approach to
  determine optimal length of k-mer.
\textit{Scientific Reports} 7:40712

\bibitem{otu2003new}
Otu HH, Sayood K. 2003.
A new sequence distance measure for phylogenetic tree construction.
\textit{Bioinformatics} 19:2122--2130

\bibitem{li2004similarity}
Li M, Chen X, Li X, Ma B, Vit{\'a}nyi PM. 2004.
The similarity metric.
\textit{IEEE Transactions on Information Theory} 50:3250--3264

\bibitem{yu2010novel}
Yu C, Liang Q, Yin C, He RL, Yau SST. 2010.
A novel construction of genome space with biological geometry.
\textit{DNA Research} 17:155--168

\bibitem{wu2001statistical}
Wu TJ, Hsieh YC, Li LA. 2001.
Statistical measures of {DNA} sequence dissimilarity under {M}arkov chain
  models of base composition.
\textit{Biometrics} 57:441--448

\bibitem{vinga2004comparative}
Vinga S, Gouveia-Oliveira R, Almeida JS. 2004.
Comparative evaluation of word composition distances for the recognition of
  {SCOP} relationships.
\textit{Bioinformatics} 20:206--215

\end{thebibliography}
\bibliographystyle{ar-style3}

%
%

\end{document}